\documentclass[5p,number,sort&compress]{elsarticle}
\usepackage{graphicx}
\usepackage{dcolumn}
\usepackage{amsmath}
\usepackage{amssymb}
\usepackage{subfigure}
\usepackage{float}
\usepackage{natbib}
\usepackage{color}

\begin{document}

\newcommand{\ket}[1]{\left| #1 \right\rangle}
\newcommand{\bra}[1]{\left\langle  #1 \right|}

\title{Detecting Majorana modes through Josephson junction ring-quantum dot hybrid architectures}
\author[mg]{Rosa Rodr\'{i}guez-Mota\corref{cor1}}
\author[uc]{Smitha Vishveshwara}
\author[mg]{T. Pereg-Barnea}
\address[mg]{Department of Physics and the Centre for Physics of Materials, McGill University, Montreal, Quebec,
Canada H3A 2T8}
\address[uc]{Department of Physics, University of Illinois at Urbana-Champaign, Urbana, Illinois 61801-3080, USA}
\cortext[cor1]{Corresponding author. \\
\textit{E-mail address}: rosarm@physics.mcgill.ca}
\date{\today}

\begin{abstract}
Unequivocal signatures of Majorana zero energy modes in condensed matter systems and manipulation of the associated electron parity states are highly sought after for fundamental reasons as well as for the prospect of topological quantum computing.
In this paper, we demonstrate that a ring of Josephson coupled topological superconducting islands threaded by magnetic flux and attached to a quantum dot acts as an excellent parity-controlled probe of Majorana mode physics.
As a function of flux threading through the ring, standard Josephson coupling yields a $\Phi_0=h/(2 e)$ periodic features corresponding to $2\pi$ phase difference periodicity. In contrast, Majorana mode assisted tunneling provides additional features with $2\Phi_0$ ($4\pi$ phase difference) periodicity, associated with single electron processes.
We find that increasing the number of islands in the ring enhances the visibility of the desired $4\pi$ periodic components in the groundstate energy. Moreover as a unique characterization tool, tuning the occupation energy of the quantum dot allows controlled groundstate parity changes in the ring, enabling a toggling between $\Phi_0$ and $2\Phi_0$ periodicity.  
\end{abstract}

\maketitle

\section{Introduction}
Majorana zero modes (MZM) have captivated condensed matter theorists and experimentalists alike of late~\citep{0034-4885-75-7-076501,RevModPhys.87.1037,0268-1242-27-12-124003,doi:10.1146/annurev-conmatphys-030212-184337} from the fundamental perspective as well as for their potential application in  topological quantum computation~\citep{1063-7869-44-10S-S29,RevModPhys.80.1083,ISI:000290150300017}.  
Progress toward the realization of MZM has been made by several theoretical proposals~\citep{PhysRevLett.105.077001,PhysRevLett.105.177002,1402-4896-2015-T164-014008,PhysRevB.88.020407} as well as experimental work~\citep{Nadj-Perge602,PhysRevB.87.241401,PhysRevLett.110.126406,Mourik1003,doi:10.1021/nl303758w,ISI:000310836700016,ISI:000346257700009,ISI:000369019100004,ISI:000311888200016,2016arXiv160309611D}. 
While most experiments involving topological superconductors present zero bias conductance peaks as evidence for the existence of MZM~\citep{Nadj-Perge602,PhysRevB.87.241401,PhysRevLett.110.126406,Mourik1003,doi:10.1021/nl303758w,ISI:000311888200016,ISI:000346257700009}, this alone can not serve as proof for their existence~\citep{PhysRevB.90.174206,PhysRevB.88.064506,PhysRevLett.109.146403,1367-2630-15-5-055019,PhysRevB.87.024515,1367-2630-14-12-125011,PhysRevLett.109.267002,PhysRevLett.109.227005,PhysRevB.88.020502,PhysRevB.89.220507,PhysRevB.86.100503,PhysRevLett.109.186802,ISI:000329315000020,2017arXiv171106256M}. 
Another manifestation of the existence of MZM is the presence of $4\pi$ periodic components in the Josephson current between two topological superconductors~\citep{1063-7869-44-10S-S29,Kwon2004,PhysRevB.79.161408,:/content/aip/journal/ltp/30/7/10.1063/1.1789931,PhysRevLett.105.077001,PhysRevLett.105.177002,Badiane2013840}.
Despite encouraging experimental evidence~\citep{ISI:000369019100004,ISI:000311888200016,2016arXiv160309611D}, interpreting the presence of $4\pi$ periodic tunneling as an unequivocal sign of MZM remains problematic for three main reasons. The first is that the $4\pi$ periodicity can only be observed when the time scale over which the phase difference in the junction changes is smaller than the time scale for quasi-particle poisoning~\citep{PhysRevB.79.161408}. 
The second problem is that the 4$\pi$ periodic components in the Josephson current are generally accompanied by other, possibly much larger, $2\pi$ periodic components. 
Finally, the presence of $4\pi$ periodic components can be caused by Andreev bound states rather than MZM~\citep{Kwon2004,2012arXiv1206.4596S,PhysRevB.95.060501}.
 
\begin{figure}[t]
\centering
	\def\svgwidth{0.4\textwidth}
	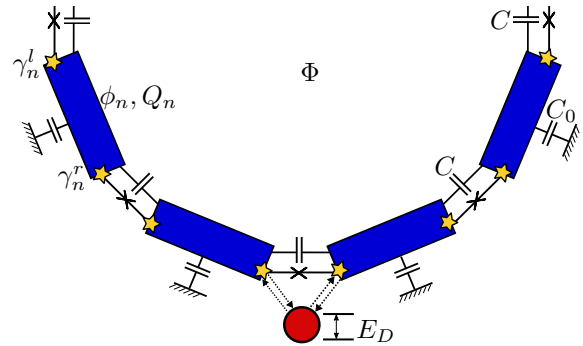
	\caption{The setup consists of a ring made of $N$ topologically superconducting islands (blue rectangles) coupled to a quantum dot (red circle) and threaded by magnetic flux $\Phi$. The islands present Majorana modes (stars) at their edges leading to single particle tunneling in addition to the usual Josephson tunneling. Electrostatic effects in the ring are modeled by self and nearest-neighbor capacitances, $C_0$ and $C$, respectively.}
	\label{fig:scheme}
\end{figure}
Our proposal to address these problems is to study the signatures of $4\pi$ periodic tunneling due to MZM in Josephson junction ring-quantum dot hybrid architectures. As will be shown, the setup we propose in this paper controls quasiparticle tunneling by tuning the capacitance of the superconducting islands and suppresses the $2\pi$ periodic Josephson contribution by connecting a number of junctions in a ring. While single particle tunneling through bound states in the junctions can only be eliminated by producing very clean junctions, our setup is able to distinguish their contribution from Majorana assisted tunneling by connecting with a quantum dot.

Here, we combine two promising MZM settings to obtain a powerful and controlled means of MZM detection--Josephson junction arrays and quantum dot geometries.
Josephson junction arrays provide a rich playground for studying the interplay between superconductivity and electrostatic repulsion~\citep{Fazio2001235}. 
These are appealing experimental systems since the relevant energy scales are relatively easy to tune, especially in one dimension~\citep{ISI:000280559300014,PhysRevB.54.R6857,PhysRevLett.81.204,Haviland2000}.
Understanding such interplay in networks of multiply-connected 1D topological superconductors is particularly important, as it is a key ingredient in proposals to detect and manipulate MZM~\citep{PhysRevB.94.115430,PhysRevB.88.035121,PhysRevLett.106.090503,PhysRevLett.106.130505,1367-2630-14-3-035019,PhysRevX.6.031016,PhysRevB.95.235305}. 
Another approach to detect and control MZM is by coupling to quantum dots and enabling single-electron hopping~\citep{PhysRevB.94.155417,PhysRevB.94.045316,PhysRevB.89.165314,Deng1557,ISI:000335642000003,1367-2630-19-1-012001,ISI:000346257700009,PhysRevB.84.201308,2017arXiv170201740C,PhysRevB.96.085418}.
Our setup builds on previous work to integrate 1D Josephson junction arrays made of topological superonconductors and quantum dots into a single architecture. 
Majorana nanowires\citep{PhysRevLett.105.077001,PhysRevLett.105.177002} provide the most natural path to physically assemble the setup studied on this work. Although more technically challenging, another possible path for physical realization could be through assembling chains of magnetic atoms on the surface of superconductors\cite{Feldman,PhysRevB.88.020407}.

The setup we study is shown in Fig.~\ref{fig:scheme}. It consists of a topological Josephson junction ring (TJJ ring) formed by $N$ topological superconducting islands threaded by magnetic flux and coupled to a quantum dot. 
Our key results are summarized in Fig.~\ref{fig:plots}. 
Assuming the absence of quasiparticle poisoning, the net parity of the ring (odd or even number of electrons) $\mathcal{P}_{TJJ}$  is conserved when it is decoupled from the quantum dot. 
Without phase fluctuations its low energy spectrum as a function of flux is a collection of parabolas centered around integer flux quanta. These parabolas corresponds to different angular momentum states for which the winding of the superconducting phase across the TJJ ring is a multiple of $2\pi$.
The contours are essentially the same as those obtained for non-topological rings with one crucial difference.  
When $\mathcal{P}_{TJJ}=1(-1)$, only parabolas which are centered around odd(even) integer flux quanta are possible.  
This is shown for $\mathcal{P}_{TJJ}=1$ in Fig.~\ref{fig:plotsm}.
Once phase fluctuations, induced by the charging energy, are included, quantum phase slips occur, creating avoided crossings in the spectrum as shown in Fig.~\ref{fig:plotpf}.  
While in the non-topological rings phase slips create a $\Phi_0$ periodic spectrum, the spectrum of the TJJ ring in the presence of phase slips is $2\Phi_0$ periodic. 
This is a consequence of parity conservation forbidding the existence of either the even or the odd parabolas. 
Upon coupling to the quantum dot, the TJJ ring can violate parity conservation by accepting or donating an electron to the dot, thus hybridizing the odd and even parity sectors and tuning the periodicity of the ring from $2\Phi_0$ to $\Phi_0$. 
The associated energy spectrum as a function of flux, measurable via persistent current, then takes on a characteristic form depending on quantum dot parameters, as shown in  Figs.~\ref{fig:TJJqd} and~\ref{fig:TJJqdpf}.

As we show in what follows, several features of this architecture together yield distinct advantages in isolating MZM physics. In contrast to a single topological junction, in the TJJ ring the effects of the $2\Phi_0$ periodic tunneling are amplified by increasing the number of islands, $N$.
Due to the charging energy of the islands, $E_0=\mathrm{e}^2/(2 C_0)$, and the occupation energy of the dot, $E_D$, there is an energy shift $\Delta E$ between the even and odd parity spectrum of the ring. 
The characteristic dependence of the energy spectrum on $\Delta E$ rules out the possibility of this effect being caused by Andreev boundstates.
A large value of the self-charging energy $E_0$ helps suppress quasi-particle poisoning arising from undesired electron and hole excitations. 
The dot's affinity to accept or donate an electron is easily controlled via applying a gate voltage and altering $E_D$. 
Tuning $\Delta E$ in this setup allows toggling between the two different TJJ ring parity sectors and thus pinpointing the effect of MZM via the associated tuning of the periodicity of the ring between $2\Phi_0$ and $\Phi_0$. 
 
\section{Topological Josephson junction (TJJ) ring}
To analyze the scenario in detail, let us begin by considering the TJJ ring in Fig.~\ref{fig:scheme} uncoupled to the quantum dot.
Each of the $N$ islands in the ring is characterized by a superconducting order parameter phase $\phi_n$ and a charge $Q_n$.
The islands' topological nature leads to two Majorana modes, $\gamma_n^l$ and $\gamma_n^r$, localized around the left and the right edge of the n$^{th}$ island. 
Neighboring islands interact through tunneling and electrostatic repulsion. 
To lowest order in the interaction, only tunneling processes that keep the superconductors in their ground state contribute.
These correspond to Josephson tunneling of pairs and Majorana assisted single electron tunneling.
The tunneling as well as the capacitance of the islands make up the TJJ ring Hamiltonian:
\begin{equation}
\begin{split}
&H_{TJJ}=H_J + H_M + H_C \\
&H_J=-\sum_n E_J \cos(\phi_{n+1}-\phi_n + \delta_\Phi) \\
&H_M = \sum_n E_M \left(c_n^\dagger c_n -\frac{1}{2}\right) \cos\left(\frac{\phi_{n+1}-\phi_n + \delta_\Phi}{2}\right)\\
&H_C=\frac{1}{2}\sum_{n,m} Q_n C_{nm}^{-1} Q_m,
\end{split}
\end{equation}
where $\phi_{n+1}-\phi_n + \delta_\Phi$ corresponds to the gauge invariant phase difference between the islands, with $\delta_\Phi=2\pi \Phi/(N\Phi_0)$. $H_J$ describes the Josephson tunneling, with amplitude $E_J$. $H_M$ describes the tunneling enabled by MZM with the energy scale $E_M$ and fermionic operators $c_n=(\gamma_n^r+i\gamma_{n+1}^l)/2$. $H_C$ describes the electrostatic repulsion with the capacitance $C_{nm} =  (C_0 + 2C) \delta_{n,m} - C \left( \delta_{n+1,m} + \delta_{n-1,m} \right)$, where $C_0$ is the self capacitance and $C$ is the neighboring capacitance. 
The TJJ ring has four relevant energy scales: $E_J$, $E_M$, and the charging energies $E_C= \mathrm{e}^2/(2C)$ and $E_0= \mathrm{e}^2/(2C_0)$.
We assume that the dominant energy scale is either $E_M$ or $E_J$, and that $E_C \ll E_0$~\citep{Fazio2001235}.
In this case, the TJJ ring is described by almost well defined superconducting condensate phases with small fluctuations controlled by $E_C$.

For $E_C=0$, the Hamiltonian of the system becomes $H_{TJJ}^{cl}=H_J + H_M + E_0 \frac{Q^2}{N}$, with $Q=\sum_n Q_n$. The superconducting phases become well-defined classical variables~\citep{PhysRevLett.89.096802,PhysRevB.87.174513}. Moreover the eigenstates of $H_{TJJ}^{cl}$ must have well defined occupations of the fermionic modes $c_n$. Since the occupation of the $c_n$ fermions is defined modulo 2~\citep{PhysRevLett.104.056402,PhysRevB.84.180502}, a given phase configuration corresponds to two distinct eigenstates of $H_{ring}^{cl}$ distinguished by their fermionic parity $\mathcal{P}_{TJJ}=(-1)^Q$ \footnote{To simplify the notation we measure the charge $Q$ in units of the electron charge $\mathrm{e}$.}. 
As shown in~\ref{sec:Proof2}, this leads to the following condition on the phases:
\begin{equation}
\begin{split}
\sum_n \theta_n  = 2\pi m \text{ with } \left\lbrace \begin{array}{ll}
m \enspace\text{even} & \quad\text{if}\enspace\mathcal{P}_{TJJ}=-1 \\
m \enspace \text{odd} & \quad \text{if}\enspace\mathcal{P}_{TJJ}=1 
\end{array} \right. ,
\end{split}
\label{eqn:loopcons}
\end{equation} 
where $\theta_n = \phi_{n+1}-\phi_n+2\pi c_n^\dagger c_n \mod 4\pi$.
The energy of a configuration of phase differences $\boldsymbol{\theta}=(\theta_1,...,\theta_N)$ can be written as $E\left(\boldsymbol{\theta}\right) = -\sum_n V \left(\theta_n + \delta_\Phi \right)$, where $V(\theta)$ is the single junction potential $V(\theta)= - E_J \cos \theta -\frac{E_M}{2} \cos \frac{\theta}{2}$.

\begin{figure}[h]
\subfigure[ $H_{TJJ}$, $\mathcal{P}_{TJJ}=1$, $E_C=0$]{
\includegraphics[width=0.23\textwidth]{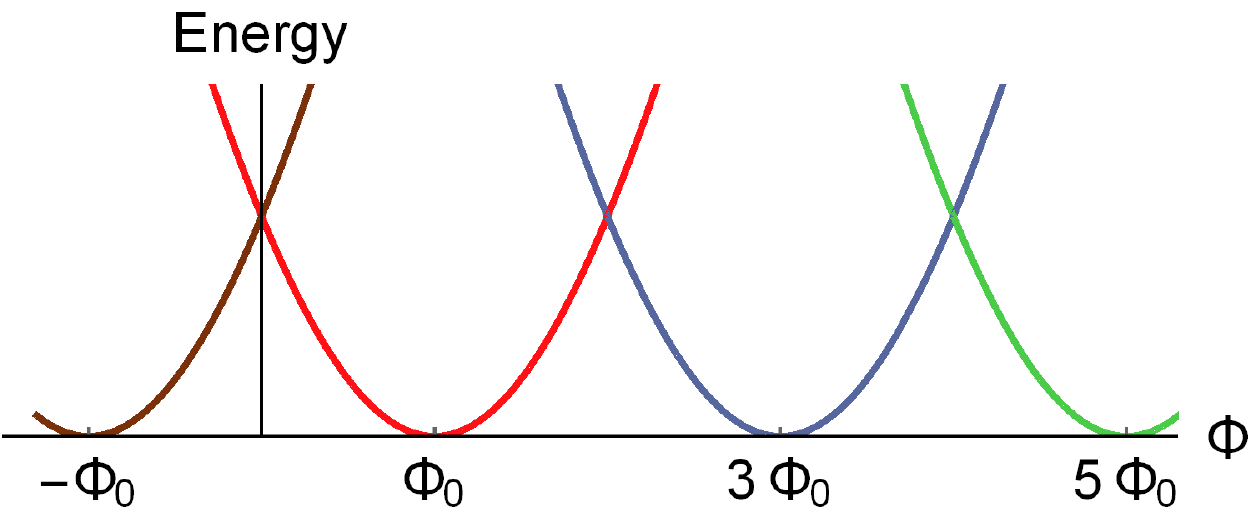}\label{fig:plotsm}}
\subfigure[ $H_{TJJ}$, $\mathcal{P}_{TJJ}=1$, $E_C>0$ ]{
\includegraphics[width=0.23\textwidth]{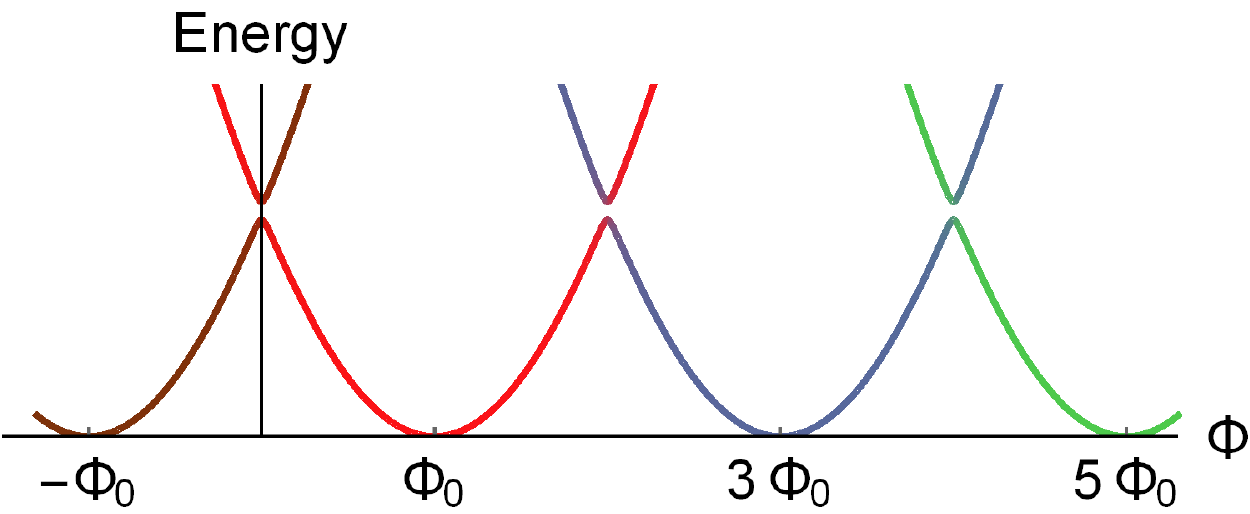}\label{fig:plotpf}}
\subfigure[ $H$, $E_C=0$ ]{
\includegraphics[width=0.23\textwidth]{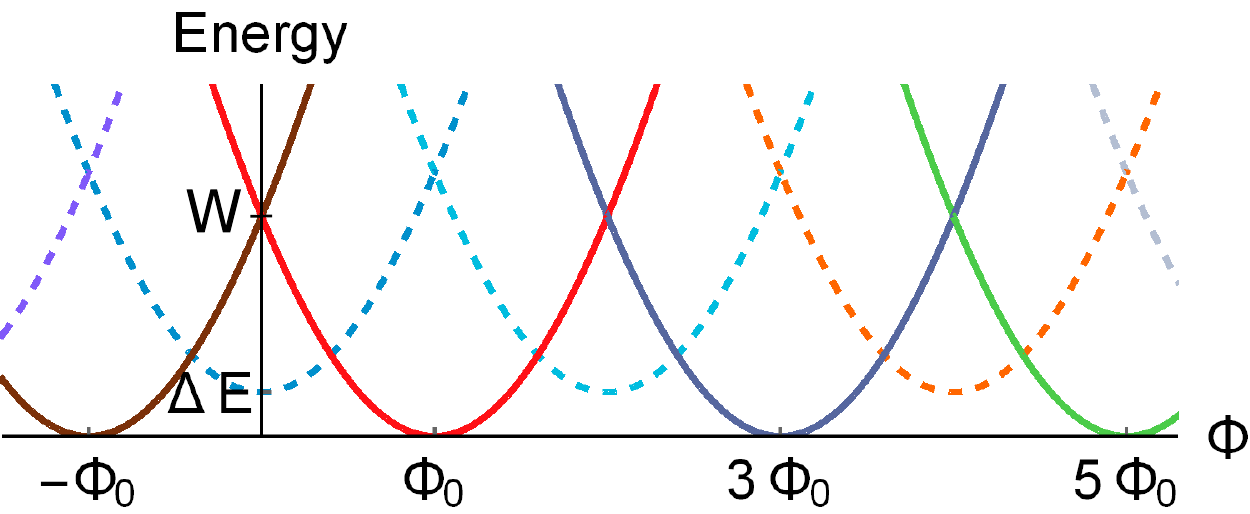}\label{fig:TJJqd}}
\subfigure[ $H$, $E_C>0$ ]{
\includegraphics[width=0.23\textwidth]{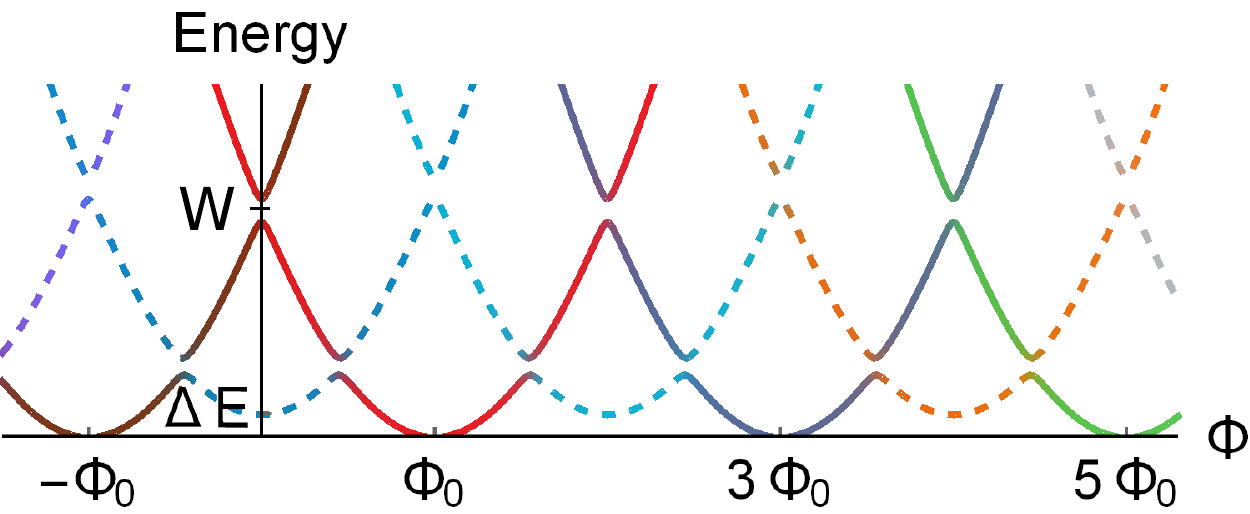}\label{fig:TJJqdpf}}
\caption{Schematic of our results for `long' TJJ rings. In this case, the $2\Phi_0$ periodic terms become dominant. \subref{fig:plotsm} Without phase fluctuations, the lowest energy bands of the even parity TJJ ring  ($\mathcal{P}_{TJJ}=1$) consist of parabolas centered around odd multiples of $\Phi_0$,  each corresponding to a different winding of the superconducting phase across the TJJ ring.
\subref{fig:plotpf} Phase fluctuations in the TJJ ring create avoided crossings making the spectrum $2\Phi_0$ periodic. The corresponding spectrum for the odd parity TJJ ring ($\mathcal{P}_{TJJ}=-1$) is that of panels \subref{fig:plotsm} and  \subref{fig:plotpf} with a $\Phi_0$ shift in the flux.
 \subref{fig:TJJqd} Once the TJJ ring is coupled to the dot, the energy spectrum includes states with $\mathcal{P}_{TJJ}=1$ (solid lines) and states with $\mathcal{P}_{TJJ}=-1$ (dashed lines). Due to charging costs, the energies of states with $\mathcal{P}_{TJJ}=-1$ and $\mathcal{P}_{TJJ}=1$ are offset by $\Delta E$. \subref{fig:TJJqdpf} Phase fluctuations lead to avoided crossings. The groundstate energy behavior depends on how $\Delta E$ compares to the bandwidth of the $\mathcal{P}_{TJJ}=1$ sector $W$.}
\label{fig:plots}
\end{figure}

The TJJ ring has a translational symmetry, i.e. the system is unchanged by circular shifts of the islands. Because of this, we expect configurations with uniform phase differences, i.e. $\theta_n=\theta$, to have the lowest energy. While this is true when $E_J=0$, for non zero $E_J$ the competition between $2\pi$ and $4\pi$ periodic tunneling may favor non uniform phase configurations. Nonetheless, we find that uniform phase configurations minimize the energy whenever 
\begin{equation}
N E_J \left( 1-\cos \frac{2\pi}{N}\right) + \frac{N E_M}{2} \left(1-\cos \frac{\pi}{N} \right)< E_M.
\label{eqn:largeN}
\end{equation}
For $N\gtrsim 6$ this condition becomes
\begin{equation}
\frac{N E_M}{\pi^2} >2E_J +\frac{ E_M}{4}.
\label{eqn:largelargeN}
\end{equation} 
As a result of the presence of $2\pi$ periodic tunneling TJJ rings exhibit local minima at even (odd) $\Phi_0$ for $\mathcal{P}_{TJJ}=1(-1)$ if condition (\ref{eqn:largeN}) is not met. Increasing $N$ reduces the role of the $2\pi$ periodic components in the lowest energy bands. For the remainder of this work, we refer to the TJJ ring as `long' if the condition (\ref{eqn:largeN}) is met and as `short' if it is not.

Since a TJJ ring with all equal junctions is a highly idealized situation, it is worth discussing how disorder in the couplings may affect the reduction of the role of $2\pi$ periodic components with increasing number of islands $N$. For $N\gtrsim 6$ and relatively small disorder the condition (\ref{eqn:largeN}) becomes
\begin{equation}
 \sum\limits_{n=1}^{N} \frac{1}{{E_J}_n + \frac{{E_M}_n}{8}} >  \frac{2\pi^2 }{ \text{min} \left( {E_M}_n \right)},
\label{eqn:largeNdisorder}
\end{equation}
where ${E_{J}}_n$ and ${E_M}_n$ are the Josephson and Majorana couplings for the $n$th junction, respectively. The above condition reduces to (\ref{eqn:largelargeN}) for even couplings. If we assume the couplings ${E_{J}}_n$ and ${E_M}_n$ to be uniformly distribution on the intervals $(E_J-\sigma_J,E_J+\sigma_J)$ and $(E_M-\sigma_M,E_M+\sigma_M)$, taking the average of  (\ref{eqn:largeNdisorder}) results in
\begin{equation}
\frac{N}{\pi^2} \left({E_M} -\sigma_M \frac{N-1}{N+1} \right) >   2{E_J}+ \frac{{E_M}}{4}.
\label{eqn:largeNdisorderAv}
\end{equation}
We conclude that some disorder in the ${E_{J}}_n$ couplings is not likely to affect our results. On the other hand, a large spread of ${E_M}_n$ couplings increases the likelihood of finding local minima on the TJJ ground-state energy. Despite this, the left hand size of (\ref{eqn:largeNdisorderAv}) grows with $N$ as long as  $\sigma_M<E_M$. Thus we conclude that the enhancement of the $4\pi$ periodic effects with increasing $N$ is stable to small disorder in the couplings.

In the following, we focus on long TJJ rings. Taking into account the constraint, Eq.~(\ref{eqn:loopcons}), the possible constant phase configurations are given by $\theta=2\pi m/N$, where $m$ is an odd(even) integer if ${P}_{TJJ}=1(-1)$. 
We label these configurations by $\ket{m}$ and their energy by $\epsilon_m= N V(2\pi(m+\Phi/\Phi_0)/N)$. These different states correspond to different angular momentum values and can be distinguished by their persistent currents.
The low-energy part of the spectrum of the states $\ket{m}$ for $\mathcal{P}_{TJJ}=1$ is shown in Fig.~\ref{fig:plotsm}. 
For $N\gtrsim6$ these states are essentially parabolas centered around $-m\Phi_0$. 

For $E_C>0$, the main types of phase fluctuations for the TJJ ring are plasmons and phase slips. Plasmons are harmonic fluctuations around the $\ket{m}$ states. They add a zero point motion energy to $\epsilon_m$. We find that plasmons in the TJJ behave similarly to plasmons in non-topological JJ rings with the plasma frequency: $\hbar\omega_p = \sqrt{8 E_J E_C + E_M E_C}$, as opposed to the non-topological frequency $\hbar\omega_p = \sqrt{8 E_J E_C}$. Phase slips lead to quantum tunneling between the $\ket{m}$ states~\citep{PhysRevLett.89.096802}, causing the avoided crossings in Fig.~\ref{fig:plotpf}. For instance, the states $\ket{m}$ and $\ket{m+2}$ are connected trough $4\pi$ phase slips. Since $H_{TJJ}$ conserves $\mathcal{P}_{TJJ}$ phase slips occur only in multiples of $4\pi$, i.e. in long TJJ rings $2\pi$ phase slips are suppressed, as in topological superconducting wires~\citep{1063-7869-44-10S-S29,PhysRevB.87.064506}.

\section{TJJ ring-quantum dot architecture}
To control the parity of the TJJ ring, we couple the ring to a quantum dot, enabling electrons to tunnel between the TJJ and dot (together referred to as TJJ+D). In the simplest case of a single electronic level available to the dot, its  Hamiltonian takes the form $H_{D}=E_{D}(d^\dagger d-\frac{1}{2})$, where $d$ and $d^\dagger$ annihilate and create an electron in the dot. 
We consider a setup where electron tunneling from the quantum dot is into MZM modes on TJJ islands 1 and $N$ with amplitudes $w_1$ and $w_N$, respectively. The Hamiltonian of the system is then $H=H_{ring}+H_D+H_{int}$, with the interaction between the TJJ ring and the dot given by:
\begin{equation}
\begin{split}
H_{int}= \frac{w_N e^{-\frac{i\phi_N}{2}}}{2} i\gamma_N^r d^\dagger + \frac{w_1e^{-\frac{i\phi_1}{2}}}{2} \gamma_1^l  d^\dagger + \mathrm{h.c.}
\end{split}
\end{equation}
Assuming that no magnetic flux is enclosed by the loop formed between the dot and the two islands, the phase difference between $w_1$ and $w_N$ is $\frac{\delta_\Phi}{2}$. The total parity is conserved in the TJJ+D system while it is not in the TJJ ring portion. 

To proceed with the TJJ+D analysis, we denote by $\ket{\boldsymbol{\theta},Q;n_d}$ a state of the system where 1) the TJJ has well defined phase differences $\boldsymbol{\theta}$ and well defined total charge Q and 2) the charge in the dot is $n_d$.
$H_{int}$ induces a $2\pi$ shift in the $Nth$ junction when moving a particle from the TJJ to the dot.
Thus, it connects the states $\ket{\boldsymbol{\theta},Q;0}$ and $\ket{\boldsymbol{\theta}-2\pi\vec{e}_N,Q-1;1}$, where $\boldsymbol{\theta}-2\pi\vec{e}_N =(\theta_1, ... ,\theta_{N-1},\theta_N-2\pi)$.
When $E_C=0$, both $\ket{\boldsymbol{\theta},Q;0}$ and $\ket{\boldsymbol{\theta}-2\pi\vec{e}_N,Q-1;1}$ are eigenstates of $H_{TJJ}+H_D$. 
As shown in~\ref{sec:proofE}, $H$ is then diagonalized by superpositions of the form 
\begin{equation}\label{eq:phi_states}
\alpha_\pm \ket{\boldsymbol{\theta},Q;0} + \beta_\pm\ket{\boldsymbol{\theta}-2\pi\vec{e}_N,Q-1;1}
\end{equation} 
with the following energies:
\begin{subequations}
\begin{equation}
\begin{split}
E_\pm (\boldsymbol{\theta}) = \sum_{n=1}^{N-1} V \left(\theta_n + \delta_\Phi \right) + V_\pm \left(\theta_N + \delta_\Phi\right),
\end{split}
\end{equation}
with,
\begin{equation}
\begin{split}
V_\pm(\theta)& = -E_J \cos \theta
 \pm \sqrt{{\left(\frac{E_M}{2} \cos\frac{\theta}{2} + \frac{\Delta E}{2}\right)}^2+w_\theta}, \\
w_\theta&= \frac{\left| w_N \right|^2 + \left| w_1 \right|^2 }{4} +\frac{\left| w_N \right| \left|w_1 \right|}{2} \cos \frac{\theta}{2}, \quad \text{and} \\
\Delta E& = E_D - E_0\left( 2Q -1\right)/N.
\end{split}
\end{equation}\label{eqn:ETJJ+D}
\end{subequations}
The offset, $\Delta E$, originates from the charging costs of the dot and the TJJ ring.

 \addtocounter{footnote}{-1} 

\begin{figure}[h]
\subfigure[]{
\includegraphics[height=0.266\textwidth, trim=8 0 4 0,clip]{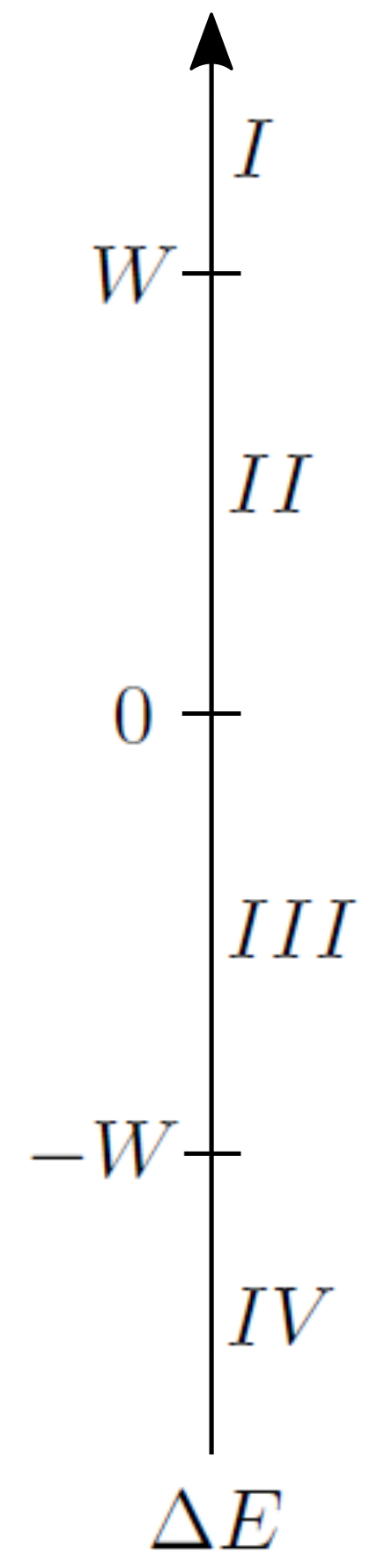}
\label{fig:pd}}
\subfigure[]{
\includegraphics[height=0.266\textwidth, trim=0 0 2 0,clip]{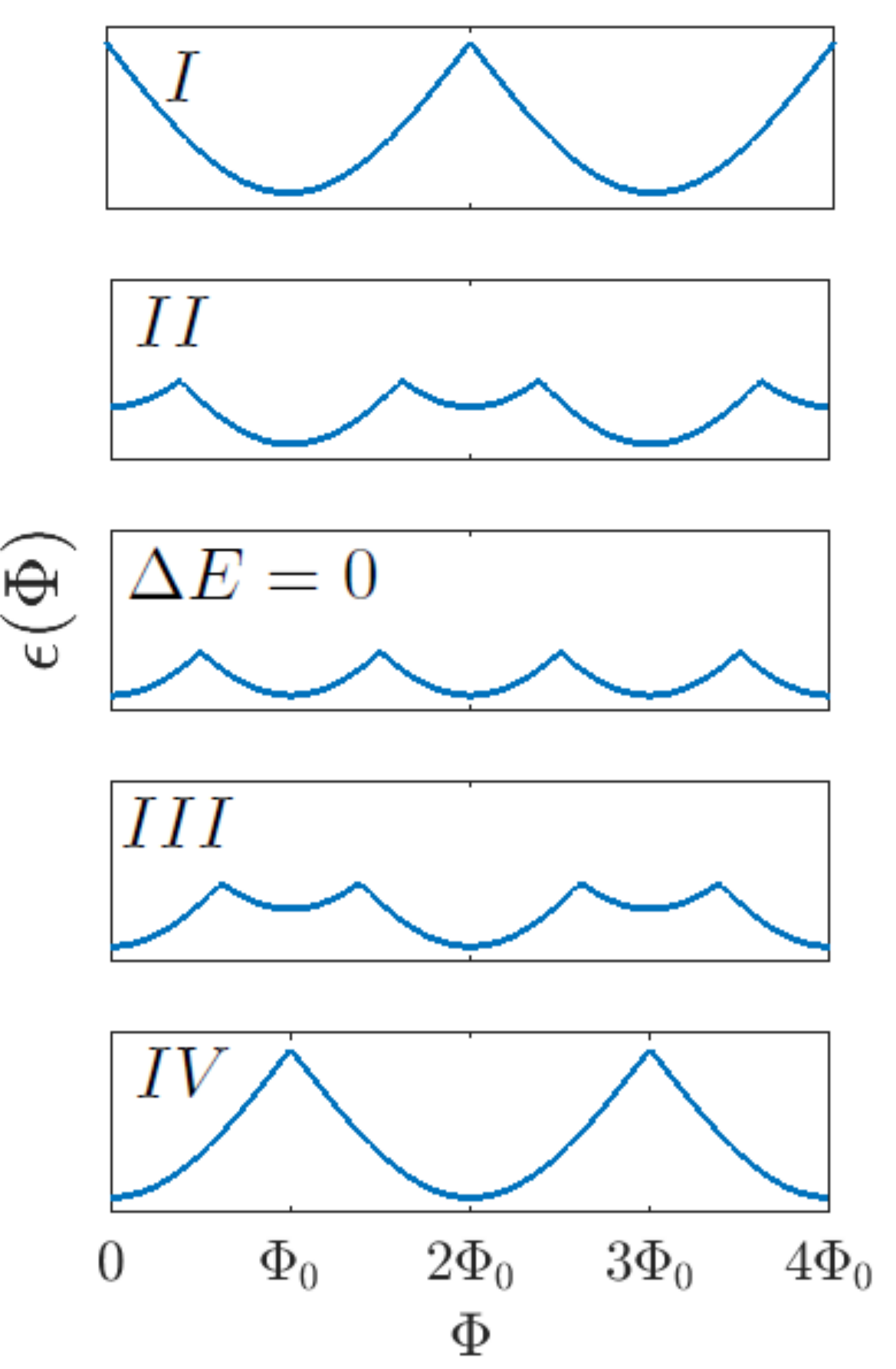}
\label{fig:energy}}
\subfigure[]{
\includegraphics[height=0.266\textwidth, trim=0 0 2 0,clip]{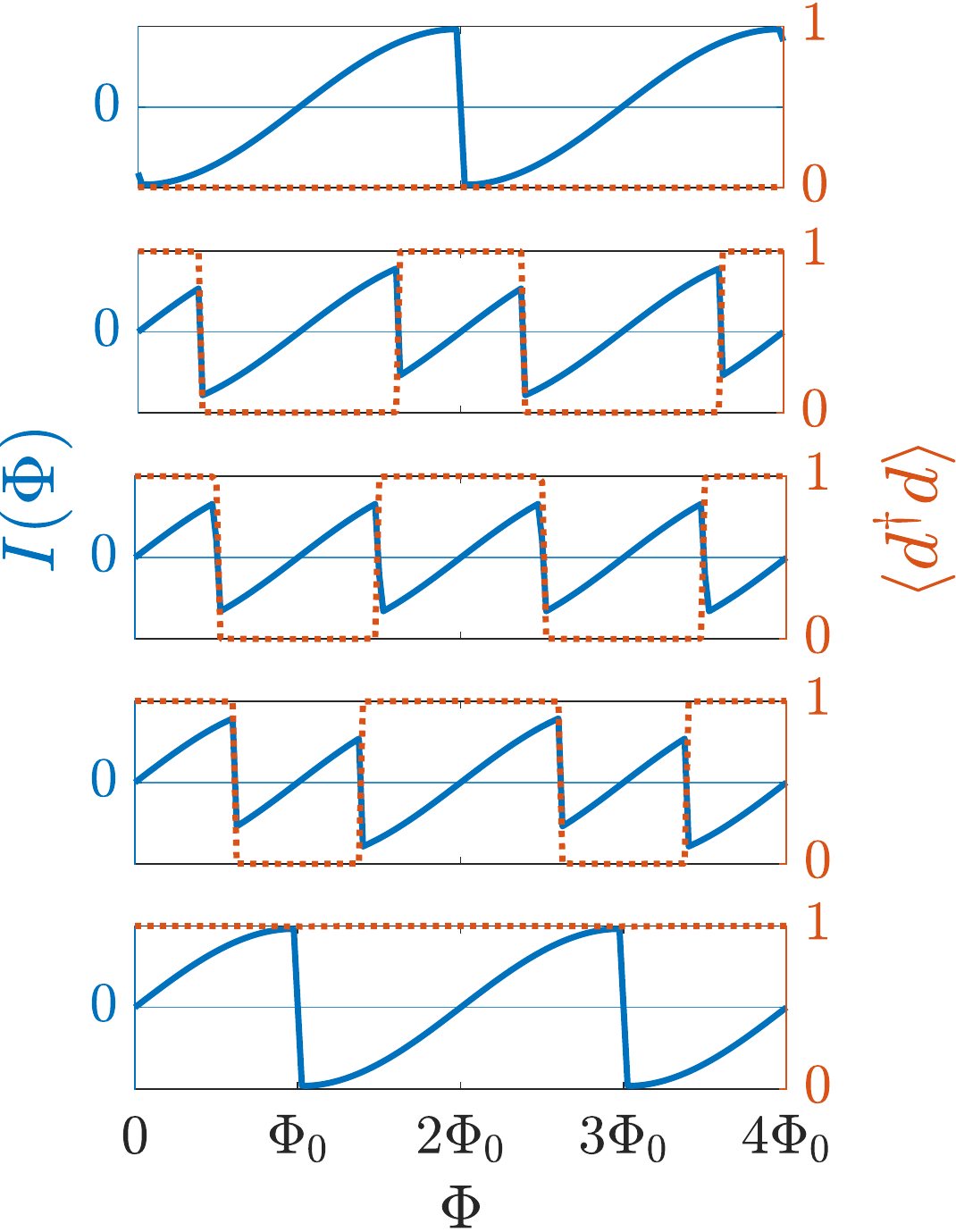}
\label{fig:current}}
\caption{The energy and current profile of the TJJ+D system in different regions of energy offset $\Delta E$ relative to the band width $W$. The different behavior provides a signature of the Majorana assisted tunneling. \subref{fig:pd} The energy offset $\Delta E$ compares to the bandwidth of the even/odd sector, $W$. \subref{fig:energy} The dependence of the groundstate energy on the magnetic flux for the TJJ+D system for the different regions in \subref{fig:pd}. \subref{fig:current} The flux dependence of the persistent current (solid blue) and the average occupation of the quantum dot (dashed red). Figures \subref{fig:energy} and \subref{fig:current} show numerical results.\protect\footnotemark}
\label{fig:numerics}
\end{figure}

\footnotetext{The results shown in Fig.~\ref{fig:numerics} were obtained through numerical simulations with the following parameters: $N=2$, $E_J=0$, $E_M=1$, $Q=100$, $E_0=0.001=10E_C$, $w_1=w_N=0.1$ and (top to bottom) $\Delta E=1.1$, $\Delta E=0.25$, $\Delta E=0$, $\Delta E=-0.25$ and $\Delta E=-1.1$.}
The TJJ+D groundstate energy, $\epsilon$, is obtained minimizing $E_- (\boldsymbol{\theta})$.
The interaction breaks the translational symmetry of the TJJ ring making the values of $\boldsymbol{\theta}$ that minimize $E_- (\boldsymbol{\theta})$ flux dependent. 
Fortunately, the TJJ+D groundstate is well approximated by flux independent states which we label $\ket{\psi_m}$. 
The states $\ket{\psi_m}$ are obtained when taking Eq.~\ref{eq:phi_states} and choosing the phase configuration of the first term to be uniform with each junction having a phase difference $2\pi m/N$ and the appropriate charge on the dot.
Furthermore, $\ket{\psi_m}$ is dominated by its component with constant phase differences in the TJJ ring, with the phase difference and occupation of the dot which match the overall parity and flux threaded.
The energies of the states $\ket{\psi_m}$, $\epsilon_m$, shown in Fig.~\ref{fig:TJJqd}, are essentially parabolas centered around even and odd multiples of $\Phi_0$, offset by $\Delta E$. 
The greatest deviation between the energies $\epsilon_m$ and $\epsilon$ is at half-integer flux values for small numbers of islands. 
Comparing the energies $\epsilon_m$ with $\epsilon$ obtained numerically for $N=2$ and $\Phi=\Phi_0/2$, we find that $\epsilon$ and the lowest $\epsilon_m$ differ by less than $0.05E_M$ for $|w_1|,|w_N|< E_M/2$. Increasing the number of islands to $N=3$ reduces such difference to less than $0.001E_M$.
The $\epsilon_m$ are then good approximations to $\epsilon$ as long as $|w_1|,|w_N| \lesssim E_M$. Further details are given in \ref{ap:fi}. 

Turning on $E_C$ leads to avoided crossings where the energies of the states $\ket{\psi_m}$ cross.
The states $\ket{\psi_m}$ and $\ket{\psi_{m\pm 1}}$ are now connected by $2\pi$ phase slips enabled by breaking the parity of the TJJ ring through the interaction with the dot. 
The behavior of the energy and that of the persistent current is then determined by where and whether the states $\ket{\psi_m}$ and $\ket{\psi_{m\pm 1}}$ cross.
This depends on how the energy offset between the even and the odd $\ket{\psi_m}$ states, $\Delta E$, compares to the bandwidth of the even (or odd) $\ket{\psi_m}$ states, $W$.
To provide a more accurate analysis, we perform numerical simulations for small island numbers.
These were done through exact diagonalization of the TJJ+D Hamiltonian limiting the charge on each island to some maximum charge $Q$. 
Examples of the different types of behavior of the energy and the persistent current obtained numerically are shown in Fig.~\ref{fig:energy} and in Fig.~\ref{fig:current}, respectively. The corresponding groundstate occupation of the dot (red line in Fig.~\ref{fig:current}) is also shown. The rapid changes in the dot groundstate occupation could be measured as peaks in the conductance as suggested by Ref.~\cite{PhysRevB.84.201308} in a similar setting.
For $\left| \Delta E \right|>W$ (regions $I$ and $IV$ in Fig.~\ref{fig:numerics}), the first energy crossing occurs between states $\ket{\psi_m}$ and $\ket{\psi_{m \pm 2}}$. In this case, the energy has global minima at either even or odd multiples of $\Phi_0$. 
On the other hand, for $\left| \Delta E \right|<W$ (regions $II$ and $III$ in Fig.~\ref{fig:numerics}), the first energy crossing occurs between states $\ket{\psi_m}$ and $\ket{\psi_{m \pm 1}}$, leading to both local and global energy minima. 

The results shown in Fig.~\ref{fig:numerics} describe the qualitative behavior of the TJJ+D architecture when the TJJ ring is long. 
For short TJJ rings, the competition between $2\pi$ and $4\pi$ periodic tunneling leads to local minima in the energy-flux relation even when $\mathcal{P}_{TJJ}$ is conserved. 
In this case, the energy of the TJJ+D system in the regions $I$ and $IV$ of Fig.~\ref{fig:numerics} would still present local minima, reducing the visibility of the transition between the two parity sectors.

The ability to tune between $2\Phi_0$ and $\Phi_0$ periodicity through controlling the occupation energy of the dot allows our setup to rule out other explanations of $2\Phi_0$ periodicity. For instance, $2\Phi_0$ periodicity may arise in small metallic or semi-conducting systems~\citep{PhysRevLett.64.2074,Bleszynski-Jayich272,PhysRevB.82.144202}. If such were the case, the $2\Phi_0$ periodicity would be unchanged by the occupation energy of the dot. If the $2\Phi_0$ periodicity was caused Andreev bound-states, the contact with a dot having small occupation energy would aid rather than suppress the $2\Phi_0$ periodicity~\cite{PhysRevB.95.060501}.

\section{Conclusions}
The proposed Josephson ring-quantum dot hybrid architecture can be realized in Josephson junction rings with Majorana nanowires~\citep{PhysRevLett.105.077001,PhysRevLett.105.177002} or with chains of magnetic atoms deposited on the surface of superconductors~\citep{Feldman,PhysRevB.88.020407}. Additionally, the TJJ ring can be understood as a coarse grained model of a 1D topological superconductor. Since the TJJ ring accounts for phase fluctuations, it could be used to shed some light into the effects of phase fluctuations, and number conservation, in topological superconductors.  Crucially, the combination of $4\pi$ periodic tunneling and the ability to manipulate the parity of the TJJ ring using the quantum dot as a knob cannot be explained through trivial Andreev bound states. Quasi-particle poisoning and $2\pi$ periodic tunneling may obscure the MZM signature. These effects can be prevented increasing the self-charging energy of the superconducting islands and increasing the number of superconducting islands, respectively.
Thus, while the Josephson junction-quantum dot hybrid architecture proposed in this paper cannot in itself enable the braiding MZMs, it can provide a solid signature of their existence. Future work would involve connecting the principles and geometry proposed here with the current scope of device capabilities in experiment.

\section*{Acknowlegments}

The authors thank D. Van Harlingen for useful discussion. This material is based upon work supported by NSERC, FQRNT (RRM, TPB), the Secretary of Public Education and the Government of Mexico (RRM) and the National Science Foundation under Grant No. 1745304 (SV).

\appendix
\section{Proof of Eqn. 2}
\label{sec:Proof2}

Due to the topological nature of each the island, for any constant phase configuration with $0 \leq \phi_n < 2\pi$ there are two superconducting ground states that can be distinguished by their fermionic parity. These groundstates will be labeled as $\ket{{\phi_n}_\mathcal{P}}$. The action of the operators $\gamma_n^{l}$ and $\gamma_n^r$ on the states $\ket{{\phi_n}_\mathcal{P}}$ is 
\begin{equation}
\begin{split}
\gamma_l \ket{{\phi_n}_\pm} = \ket{{\phi_n}_\mp} \\
i \gamma_r \ket{{\phi_n}_\pm} = \mp \ket{{\phi_n}_\mp}.
\end{split}
\label{eqn:Majoranaoperators}
\end{equation}

The Majorana operators associated with the superconducting island $n$ are given by
\begin{equation}
\begin{split}
\gamma_n^l = \int_{x \in n} \limits dx \left( e^{-\frac{i\phi}{2}} f_n^l(x) \psi^\dagger(x) +  e^{\frac{i\phi}{2}} {f_n^l(x)}^\ast \psi(x) \right) \\
\gamma_n^r = \int_{x \in n}\limits dx \left( i e^{-\frac{i\phi}{2}} f_n^r(x)\psi^\dagger(x) - i e^{\frac{i\phi}{2}} {f_n^r(x)}^\ast \psi(x)\right), 
\end{split}
\label{eqn:Majoranas}
\end{equation}
with $f_n^{l(r)}(x)$ a function localized around the left (right) edge of the $n$ island and $\psi(x)$ the field operator. Under the gauge transformation $\phi_n\rightarrow\phi_n + 2\pi$ \footnote{Note that the gauge transformation $\phi_n\rightarrow\phi_n + 2\pi$, which is a change in how we are looking at the system, differs from changing the phase $\phi_n$ by $2\pi$ adiabatically, which is a physical change in the system.}, the operators $\gamma_n^{l(r)}$ pick up a minus sign resulting in $c_n \rightarrow -c_n^\dagger$ and $c_{n-1}\rightarrow c_{n-1}^\dagger$. This implies that the occupation of the $c_n$ fermions is defined modulo 2 and care must be taken to avoid over-counting the states in the Hilbert space~\citep{PhysRevLett.104.056402}.

Following Ref.~\citep{PhysRevB.81.134435} we define the following $N-1$ independent variables
\begin{equation}
\theta_n = \phi_{n+1} - \phi_{n} +2\pi c_n^\dagger c_n \quad \text{mod} \quad 4\pi,
\label{eqn:thetas}
\end{equation}
for $n=1,...,N-1$, which are invariant under $\phi_n\rightarrow\phi_n + 2\pi$. Writing $H_J$ and $H_M$ in terms of the $\theta_n$s results in  
\begin{equation}
\begin{split}
&H_M = - \sum_{n=1}^{N-1} \frac{E_M}{2} \cos \left( \frac{\theta_n + \delta_\Phi}{2} \right) \\
&-  \frac{E_M}{2} \cos \left( \frac{-\sum_{n=1}^{N-1} \theta_n - 2\pi \sum_{n=1}^N c_n^\dagger c_n+ \delta_\Phi}{2} \right)\\
&H_J = - \sum_{n=1}^{N-1} E_J \cos \left( \theta_n + \delta_\Phi \right)
\\&-E_J \cos \left( -\sum_{n=1}^{N-1} \theta_n  + \delta_\Phi \right) , 
\end{split}
\end{equation}

The operators $\theta_n$ defined by Eqn. (\ref{eqn:thetas}) are not enough to determine the state of the TJJ since the variables $\phi_1$ and the $(-1)^{\sum_{n=1}^N c_n^\dagger c_n}$ are independent of them. To address this we define the $\theta_0$ as
\begin{equation}
e^{\frac{i\theta_0}{2}} = \gamma_1^l e^{\frac{i\phi_1}{2}}.
\end{equation}
Under the above definition $\theta_0$ remains invariant when $\phi_1\rightarrow \phi_1+2\pi$, and we have $[\theta_n,\theta_k]=0$ for all $n,k=0,...,N-1$. The operator $\theta_0$ obeys the following commutation relation with the total charge $Q=\sum_n Q_n$:
\begin{equation}
\left[ Q ,e^{\frac{i\theta_0}{2}}\right] = e^{\frac{i\theta_0}{2}}.
\end{equation}
The fact that $\theta_0$ does not appear in $H_M$ and $H_J$ indicates that both $H_M$ and $H_J$ conserve the total charge $Q$ of the TJJ. Additionally for $n=1,...,N-1$ we also have 
\begin{equation}
\left[\frac{\theta_n}{2}, Q \right] = 0 \quad \text{and} \quad
\left[\frac{\theta_n}{2}, Q_k \right] = i \left(\delta_{n+1,k} - \delta_{n,k}\right), 
\end{equation}
hence it is possible to describe the state of the TJJ using either the states $\ket{\theta_0,\theta_1,...,\theta_{N-1}}$ or the states $\ket{Q,\theta_1,...,\theta_{N-1}}$. In the following we will use the later since $[H_{TJJ},Q]=0$. 

To make the TJJ ring translational symmetry evident, it is convenient to rewrite $H_M$ and $H_J$ in terms of $N$ constrained phase differences.
This results in
\begin{equation}
\begin{split}
H_M =& - \sum \frac{E_M}{2} \cos \left( \frac{\theta_n + \delta_\Phi}{2} \right)\\
H_J =& -\sum_n E_J \cos \left( \theta_n + \delta_\Phi \right), 
\end{split}
\end{equation}
with the constraint
\begin{equation}
\begin{split}
\sum_n \theta_n  =\left\lbrace \begin{array}{ccc}
4\pi m  &\text{if} & (-1)^{\sum_n c_n^\dagger c_n}=1 \\
2\pi (2 m+ 1) &\text{if} & (-1)^{\sum_n c_n^\dagger c_n}=-1
\end{array} \right. .
\end{split}
\label{eqn:constraint}
\end{equation} 

It is also possible to relate $(-1)^{\sum_n c_n^\dagger c_n}$ to $Q$ by noting that $(-1)^{Q_n} = i \gamma_n^r \gamma_n^l$ and $(-1)^{c_n^\dagger c_n} = i \gamma_{n+1}^l \gamma_n^r$. The relation between $(-1)^{\sum_n c_n^\dagger c_n}$ and $Q$ is then
\begin{equation}
\begin{split}
(-1)^{\sum_n c_n^\dagger c_n} = & \prod_{n=N}^1 i \gamma_{n+1}^l \gamma_n^r =  \gamma_1^l \left( \prod_{n=N}^2 i\gamma_n^r \gamma_n^l \right) i \gamma_1^r \\
= &- \prod_{n=N}^1 i\gamma_n^r \gamma_n^l = -(-1)^Q.  
\end{split}
\label{eqn:parity}
\end{equation}
Combining Eqns. (\ref{eqn:constraint}) and (\ref{eqn:parity}) leads to Eqn. (\ref{eqn:loopcons}). 

We will use $\ket{\boldsymbol{\theta}}_{Q}$ to denote the state with charge $Q$ and phase differences given by $\boldsymbol{\theta} = (\theta_1,...,\theta_N)$. 

\section{Quantifying the decrease of the $2\pi$ periodic tunneling contribution and its stability against junction disorder}

In the main text, we showed that local minima in the ground-state energy vs flux relation of the TJJ can be removed by increasing the number of islands in the TJJ. Since the local minima arise due to the contribution of $2\pi$ periodic tunneling, we used this fact to argue that increasing $N$ reduces the role of $2\pi$ periodic terms. In this appendix, we provide an additional way to quantify such decrease and use it to study the stability of this effect with respect to disorder.

The energy of the TJJ ring $E(\Phi)$ can be written as a Fourier series:
\begin{equation}
E(\Phi) = \sum_{n=0}^{\infty} E_n \cos(\pi n \Phi).
\end{equation}
Using such decomposition, we can quantify the role of $2\pi$ periodic terms on the energy as
\begin{equation}
r = \frac{ \sum_{n=1}^{\infty} |E_{2n}|^2}{\sum_{n=1}^\infty |E_{n}|^2}.
\end{equation}
If only $\Phi_0$ periodic terms are present in the energy vs flux relation, i.e.  $E_M \rightarrow 0$, then $r=1$. 

\begin{figure}[h]
\center\subfigure[]{
\includegraphics[width=0.232\textwidth]{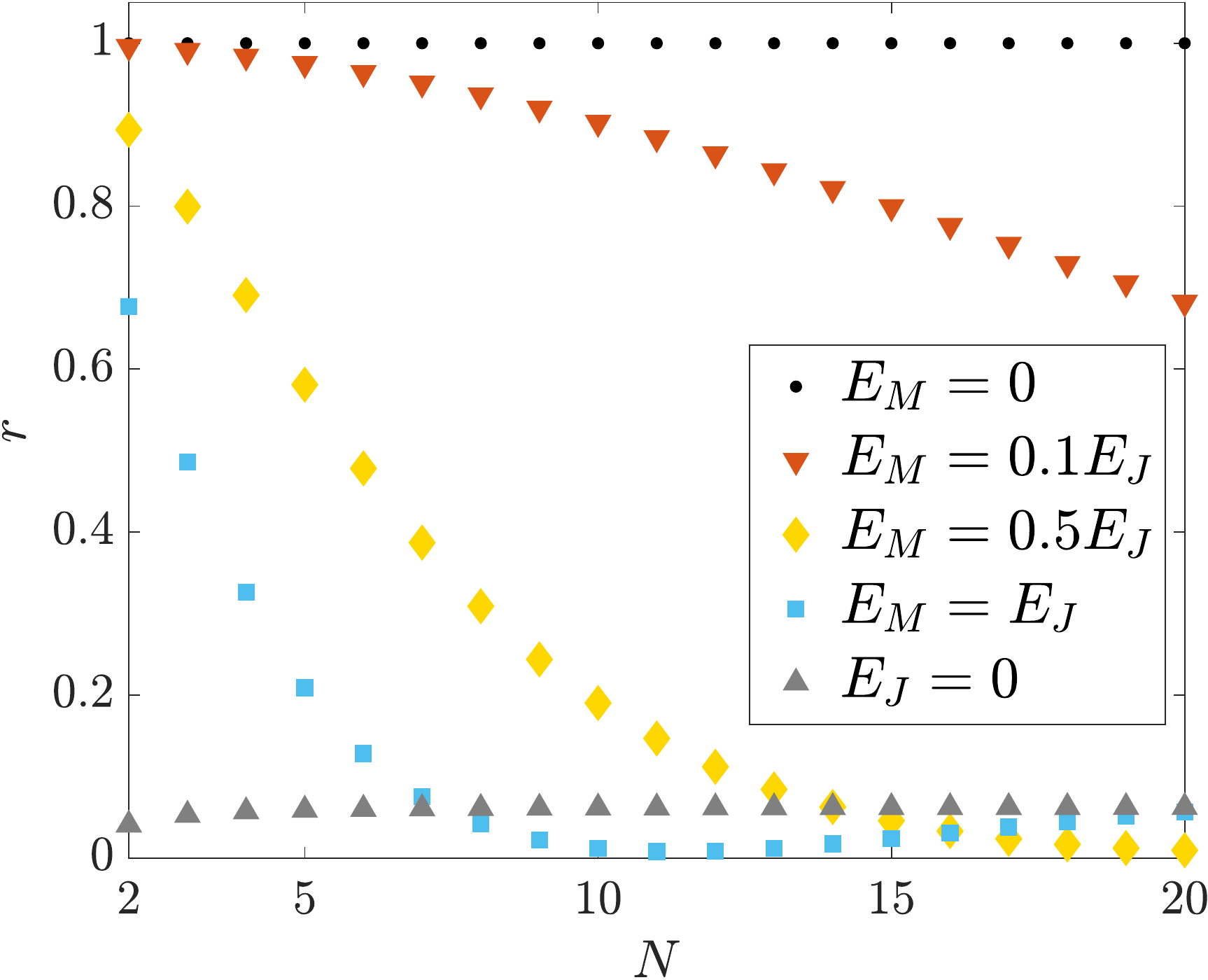}\label{fig:r}}
\subfigure[]{
\includegraphics[width=0.232\textwidth]{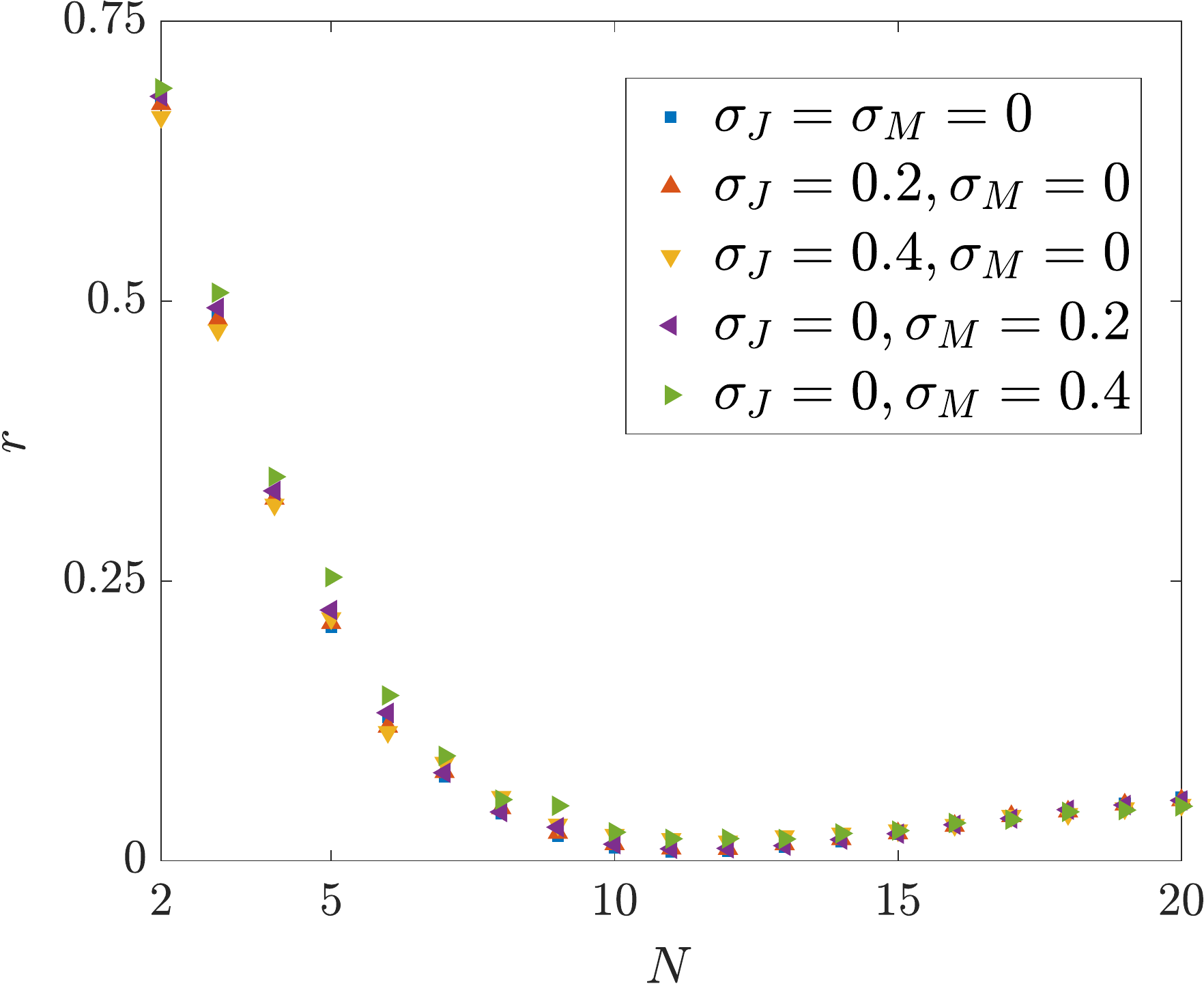}\label{fig:rdisorder}}
\caption{\subref{fig:r} Strength of the $2\pi$ periodic contribution to the ground-state energy as a function of $N$ for different rations of $E_J/E_M$. \subref{fig:rdisorder} Average strength of the $2\pi$ periodic contribution to the ground-state energy as a function of $N$ for $E_J=E_M=1$ and different amounts of disorder.}
\end{figure}

Fig. \ref{fig:r} shows $r$ as a function of the number of junctions in the ring for different rations of $E_J$ with respect $E_M$. The results where obtained minimizing the classical energy vs. flux relation of the TJJ ring numerically. As expected, $r=1$ when $E_M=0$. On the contrary, $\Phi_0$ periodic components do not fully disappear when the Cooper pair tunneling is absent, i.e. $E_J=0$. This is due to shape of the ground-state energy dependence on the flux for $E_C=0$, which is non-sinusoidal (see Fig. \ref{fig:plotsm}). Nonetheless, $r$ provides a measure for the effects of the $2\pi$ periodicity on the ground-state energy. For $E_M=E_J$ (blue squares) $r$ decreases with $N$ at first,  $r$ starts increasing after it goes below the value of $r(E_J=0)$ (gray up triangles) and then it continues to approach this value. This result agrees with our claim that the groundstate-energy dispersion for `long' TJJ rings resembles that of rings with no $2\pi$ periodic tunneling, i.e. $E_J=0$. 
The $r$ dependence on $N$ for $E_M=0.1E_J$ (down red triangles) and $E_M=0.5E_J$ (yellow diamonds) seem to follow a similar trend, but the range of $N$ in Fig. \ref{fig:r} is not large enough to appreciate the full behavior. 

Figure \ref{fig:rdisorder} shows the behavior of $r$ with respect to $N$ for $E_J=E_M=1$ and different values of disorder. To obtain this figure, we calculated the average of $r$ considering that the Josephson and Majorana couplings of the islands uniformly distributed on $(E_J-\sigma_J,E_J+\sigma_J)$ and $(E_M-\sigma_M,E_M+\sigma_M)$, respectively. In Fig.~\ref{fig:rdisorder} we see that the qualitative behavior of $r$ is unchanged by disorder in Josephson and Majorana couplings. We also find that for $N$ up to 10, disorder in the Majorana hybridization energy, increases $r$. This is in agreement with the effects of disorder stated in the main text: the role of $2\pi$ periodic contributions is relatively insensitive to disorder in the Josephson couplings, on the other hand disorder on the Majorana hybridization energy increases the role of $2\pi$ periodic contributions overall. The fact that the role of $2\pi$ periodic contributions is decreased by increasing the number of islands $N$, is insensitive to relatively small disorder on both types of tunneling. 

\section{Proof of Eqn. \ref{eqn:ETJJ+D}}
\label{sec:proofE}

Here we obtain the energies of the TJJ+D system for $E_C=0$, described by $H_{TJJ}^{cl} + H_D + H_{int}$. We start by writing $H_{int}$ in terms of the operators defined in the previous section:
\begin{equation}
\begin{split}
H_{int}= &\frac{w_N e^{-\frac{i\phi_N}{2}}}{2} i\gamma_N^r d^\dagger + \frac{w_1e^{-\frac{i\phi_1}{2}}}{2} \gamma_1^l  d^\dagger + \mathrm{h.c.}
\\= &\left[- \frac{w_N}{2} e^{\frac{i}{2} \sum\limits_{n=1}^{N-1} \theta_n} (-1)^Q + \frac{w_1}{2} \right] e^{\frac{-i\theta_0}{2}}  d^\dagger
\\ &+ \mathrm{h.c.}.
\end{split}
\end{equation}

From the above equation we obtain that $H_{int}$ connects the states $\ket{Q,\theta_1,...,\theta_{N-1}}$ and $d^\dagger \ket{Q-1,\theta_1,...,\theta_{N-1}}$ as follows:
\begin{equation}
\begin{split}
&H_{int} \ket{Q,\theta_1,...,\theta_{N-1}} = - t  d^\dagger \ket{Q-1,\theta_1,...,\theta_{N-1}} \\
&H_{int} d^\dagger \ket{Q-1,\theta_1,...,\theta_{N-1}}= - t^\ast  \ket{Q,\theta_1,...,\theta_{N-1}} \\ &\text{with} \quad t = \frac{1}{2} \left(w_N e^{\frac{i}{2} \sum_{n=1}^{N-1} \theta_n} (-1)^Q + w_1 \right).
\end{split}
\end{equation}
Alternatively, we can write
\begin{equation}
\begin{split}
H_{int} \ket{\boldsymbol{\theta}}_{Q} = \text{-} \left[\frac{w_N}{2} e^{\frac{-i\theta_N}{2}} + \frac{w_1}{2} \right] d^\dagger \ket{\boldsymbol{\theta}-2\pi\vec{e}_N}_{Q-1} \\
H_{int} d^\dagger \ket{\boldsymbol{\theta}-2\pi\vec{e}_N}_{Q-1} = \text{-} \left[\frac{w_N^\ast}{2} e^{\frac{i\theta_N}{2}} + \frac{w_1^\ast}{2} \right] \ket{\boldsymbol{\theta}}_{Q}.
\end{split}
\end{equation}

The states $\ket{\boldsymbol{\theta}}_{Q}$ and $ d^\dagger \ket{\boldsymbol{\theta}-2\pi\vec{e}_N}_{Q-1} $ are eigenstates of $H_{ring}+H_d$ with 
\begin{equation}
\begin{split}
\left(H_{TJJ}+H_D\right) &\ket{\boldsymbol{\theta}}_{Q} = \left[E(\boldsymbol{\theta})+\frac{E_0 Q^2}{N} -\frac{E_D}{2} \right] \ket{\boldsymbol{\theta}}_{Q} \\
\left(H_{TJJ}+H_D\right)&d^\dagger \ket{\boldsymbol{\theta}-2\pi\vec{e}_N}_{Q-1}  = \left[ E(\boldsymbol{\theta}-2\pi\vec{e}_N)  \right.
\\  + \frac{E_D}{2} + &\left.\frac{E_0 (Q-1)^2}{N}\right] d^\dagger \ket{\boldsymbol{\theta}-2\pi\vec{e}_N}_{Q-1}
\end{split}
\end{equation}
where $E\left(\boldsymbol{\theta}\right) = -\sum_n V \left(\theta_n + \delta_\Phi \right)$,  $V(\theta)= - E_J \cos \theta -\frac{E_M}{2} \cos \frac{\theta}{2}$.

Then $H=H_{TJJ}+H_D+H_C$ is diagonalized by states of the form $\alpha_{\pm}  \ket{\boldsymbol{\theta}}_{Q} + \beta_{\pm}  d^\dagger \ket{\boldsymbol{\theta}-2\pi\vec{e}_N}_{Q-1}$ with energies $E_{\pm} \left(\boldsymbol{\theta} \right) $ given by Eqns. (5) and (6) of the main text. 

\section{Numerical Simulations.}

In order to simulate the system numerically, it is convenient to describe the system in terms of charges rather than phases. For simplicity, we will focus on the case $N=2$. We want to find out the action of $H=H_C+ H_J + H_M + H_D + H_{int}$ on a state with well defined charges on the islands and the dot, i.e., $\ket{Q_1,Q_2,d}$. States with well defined charge are eigenstates of $H_C$ and $H_D$:
\begin{equation}
\begin{split}
( H_C  + &H_D)\ket{Q_1,Q_2,0} = \\
&\left(\frac{\mathrm{e}^2}{2} \sum_{n,m=1}^2 Q_n C_{nm}^{-1} Q_m -\frac{E_D}{2}\right) \ket{Q_1,Q_2,0} \\
( H_C  + &H_D)\ket{Q_1,Q_2,1} = 
\\ &\left(\frac{\mathrm{e}^2}{2} \sum_{n,m=1}^2 Q_n C_{nm}^{-1} Q_m + \frac{E_D}{2}\right) \ket{Q_1,Q_2,1}.
\end{split}
\label{eqn:HC}
\end{equation}

Now we proceed to find the effect of the $H_J$, $H_M$ and $H_{int}$ on the constant charge states. In order to do this, we first note that for the $n$th superconducting island the constant charge state $\ket{Q_n}$ can be constructed in terms of the states $\ket{{\phi_n}_\mathcal{P}}$:
\begin{equation}
\ket{Q_n} = \frac{1}{2\pi} \int_0^{2\pi} d\phi e^{i \phi_n \frac{Q_n}{2}} \ket{\phi_\mathcal{P}},\text{ with } \mathcal{P}=(-1)^{Q_n}. 
\label{eqn:chargestates}
\end{equation}

Using Equations  \ref{eqn:Majoranaoperators} and \ref{eqn:chargestates} we can obtain the effect of the operators  $e^{\pm \frac{i \phi_n}{2}} \gamma_n^{r(l)}$ on a state of the island $n$ with well defined charge:
\begin{equation}
\begin{split}
e^{\pm \frac{i\phi_n}{2}} \gamma_n^{l} \ket{Q_n} =& \ket{Q_n \pm 1} \\
e^{\pm \frac{i\phi_n}{2}} i \gamma_n^{r} \ket{Q_n} =& - (-1)^{Q_n} \ket{Q_n \pm 1}. 
\end{split}
\end{equation}

Hence, we can write the states $\ket{Q_1,Q_2,d}$ as follows:
\begin{equation}
\begin{split}
\ket{Q_1,Q_2,d} = \left( e^{\frac{i \phi_1}{2}} \gamma_1^l \right)^{Q_1}  \left( e^{\frac{i \phi_2}{2}} \gamma_2^l \right)^{Q_2}  (d^\dagger)^d \ket{0}
\end{split}
\end{equation}

Using the above definition we find the action of $H_M$, $H_J$ and $H_{int}$ on the states $\ket{Q_1,Q_2,d}$:
\begin{equation}
\begin{split}
& H_M \ket{Q_1,Q_2,d} = \frac{E_M}{4}\times \\
&\left[ \left( (-1)^{Q_1+Q_2} e^{-\frac{i\delta_\Phi}{2}} - e^{\frac{i\delta_\Phi}{2}} \right)
  \ket{Q_1-1,Q_2+1,d}+ \right. \\
&\left. \left( (-1)^{Q_1+Q_2} e^{\frac{i\delta_\Phi}{2}} - e^{-\frac{i\delta_\Phi}{2}} \right)
\ket{Q_1+1,Q_2-1,d} \right],
\end{split}
\label{eqn:HM}
\end{equation}
\begin{equation}
\begin{split}
 H_J \ket{Q_1,Q_2,d} = -E_J \cos \delta_\Phi   \ket{Q_1-2,Q_2+2,d} \\
 -E_J \cos \delta_\Phi  \ket{Q_1+2,Q_2-2,d},
\end{split}
\label{eqn:HJ}
\end{equation}
and
\begin{equation}
\begin{split}
H_{int} \ket{Q_1,Q_2,0} = - \frac{|w_2|}{2} e^{\frac{i\delta_\Phi}{4}} \ket{Q_1,Q_2-1,1}  +  \\
 (-1)^{Q_1+Q_2} \frac{|w_1|}{2}e^{-\frac{i\delta_\Phi}{4}} \ket{Q_1-1,Q_2,1} \\
H_{int} \ket{Q_1,Q_2,1} = - \frac{|w_2|}{2} e^{\frac{i\delta_\Phi}{4}} \ket{Q_1,Q_2+1,0}  +  \\
 (-1)^{Q_1+Q_2} \frac{|w_1|}{2}e^{-\frac{i\delta_\Phi}{4}} \ket{Q_1+1,Q_2,0}.
\end{split}
\label{eqn:Hint}
\end{equation}

Since $Q_1 + Q_2 + n_d = Q$ is conserved by the Hamiltonian, we can write the Hamiltonian for a given $Q$ sector:
\begin{equation}
\begin{split}
H = &\sum_{d,d^\prime=0}^1 \sum_{Q_1,Q_1^\prime=0}^{Q-d,Q-d^\prime} H_{Q_1,d}^{Q_1^\prime,{d}^\prime}
\times \\
&\ket{Q_1,Q_1-Q-d,d} \bra{Q_1^\prime,Q_1^\prime-Q-d^\prime,d^\prime}
\end{split}
\end{equation}
where $ H_{Q_1,d}^{Q_1^\prime,{d}^\prime}$ is the matrix element between the states 
$\ket{Q_1,Q_1-Q-d,d}$ and $\ket{Q_1^\prime,Q_1^\prime-Q-d^\prime,d^\prime}$ and can be obtained from Eqns. (\ref{eqn:HC}), (\ref{eqn:HM}), (\ref{eqn:HM}) and (\ref{eqn:Hint}). The numeric results shown in the main text were obtained from the above Hamiltonian using exact diagonalization. 

The above description can be readily extended to an arbitrary number of islands $N$, as the action of $H$ on a state $\ket{Q_1,...,Q_N,d}$ can be found by considering
\begin{equation}
\begin{split}
\ket{Q_1,...,Q_N,d} = \left( e^{\frac{i \phi_1}{2}} \gamma_1^l \right)^{Q_1}&   ... \left( e^{\frac{i \phi_N}{2}} \gamma_N^l \right)^{Q_N} \times 
\\ & (d^\dagger)^d \ket{0} . 
\end{split}
\end{equation}

\section{TJJ+D ground-state energy approximation.}
\label{ap:fi}

In the main text it was argued that the TJJ+D groundstate energy $\epsilon$ was well approximated by the energies $\epsilon_m$ of flux independent states $\ket{\psi_m}$. It was also argued that such approximation works best a) close to integer flux quantum and b) when we increase the number of islands. Here we provide some details to support such arguments. First, we note that the reason the approximation of $\epsilon \approx \text{min} (\epsilon_m)$ works best close to integer flux quantum is that the state $\ket{\psi_m}$ corresponds to the ground-state of the system when $\Phi=-m\Phi_0$. 

\begin{figure}[h]
\subfigure[]{
\includegraphics[width=0.225\textwidth]{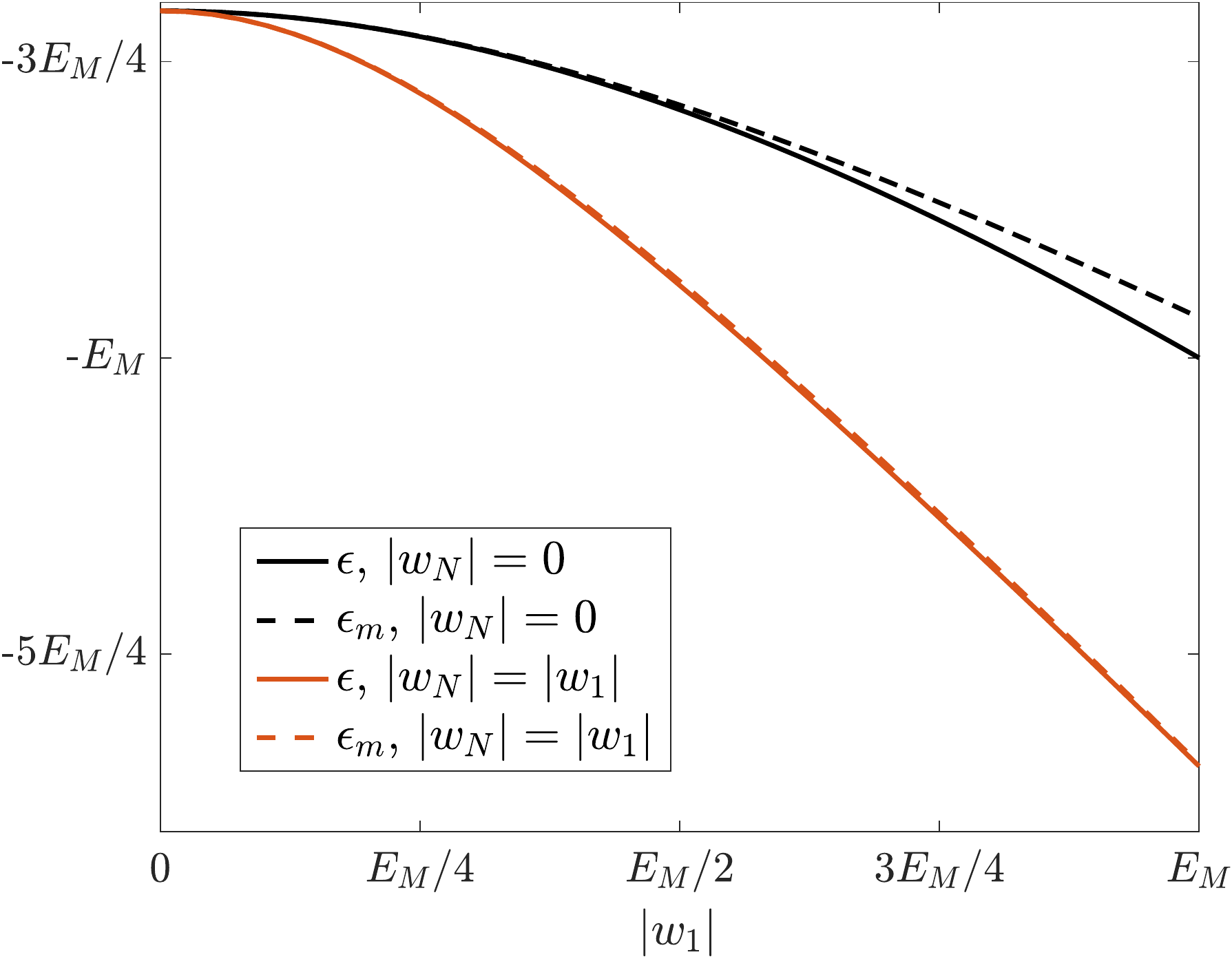}
\label{fig:N2}}
\subfigure[]{
\includegraphics[width=0.225\textwidth]{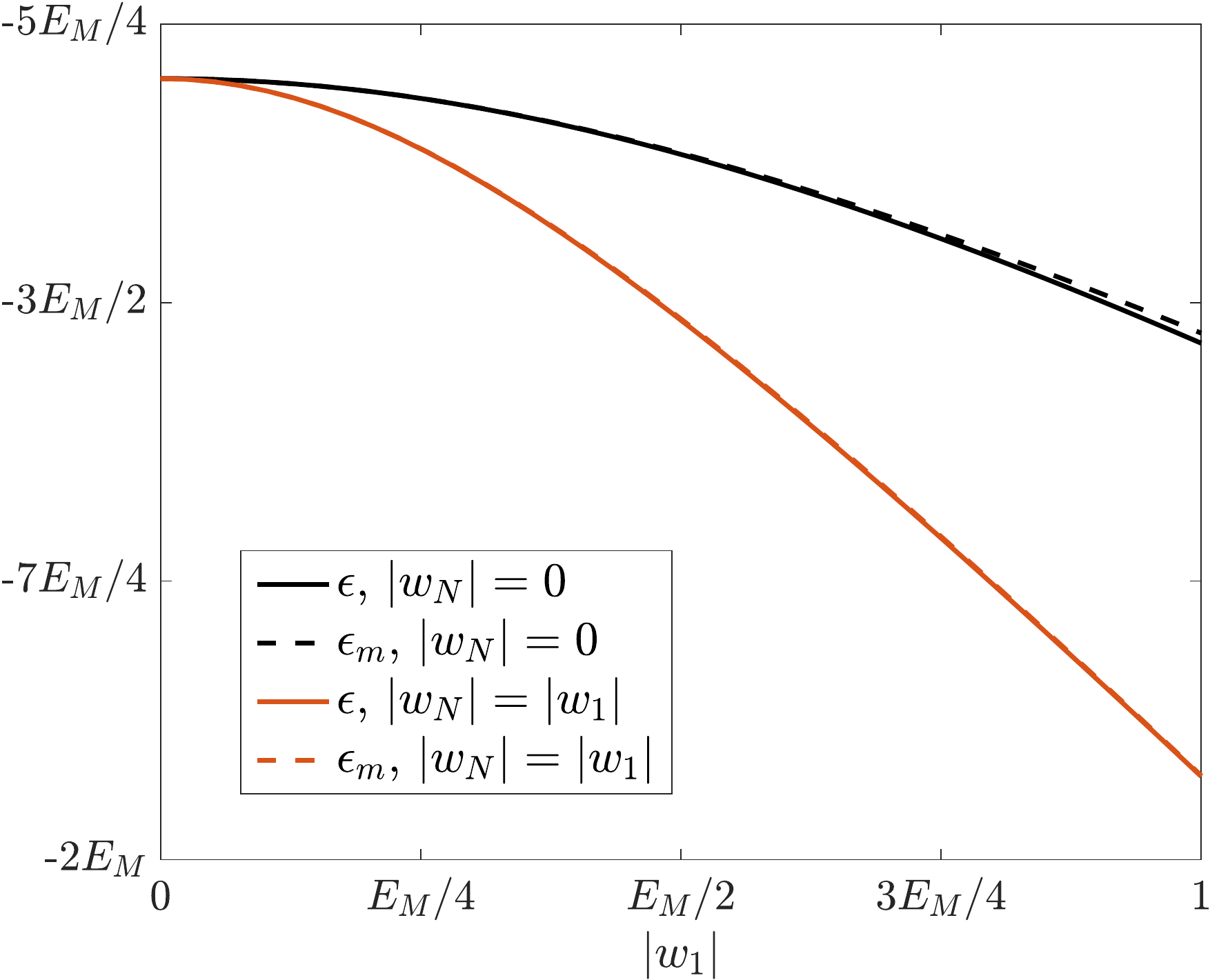}
\label{fig:N3}}
\caption{The ground-state energy of the TJJ+D, $\epsilon$ (solid lines), and $\epsilon_m$ with $m=-1$ (dashed lines) at $\Phi=\Phi_0/2$ are shown for $E_J,E_C=0$ and $N=2$ in panel \subref{fig:N2}, and $N=3$ in panel \subref{fig:N3}.}
\label{fig:approx}
\end{figure}

On the other hand, the approximation improves when $N$ increases since when the flux can be distributed in more junctions the ground-state configurations for different flux values are separated by smaller phase differences. Fig. \ref{fig:approx} shows how the considerable improvement in the approximation obtain by increasing $N$ from $N=2$ to $N=3$. The ground-state energy $\epsilon$ in Fig. \ref{fig:approx} was obtained minimizing $E_- (\boldsymbol{\theta})$ with respect the phase differences vector $\boldsymbol{\theta}$ numerically.

\bibliographystyle{apsrev}
\bibliography{biblio}

\begin{thebibliography}{72}
\expandafter\ifx\csname natexlab\endcsname\relax\def\natexlab#1{#1}\fi
\expandafter\ifx\csname bibnamefont\endcsname\relax
  \def\bibnamefont#1{#1}\fi
\expandafter\ifx\csname bibfnamefont\endcsname\relax
  \def\bibfnamefont#1{#1}\fi
\expandafter\ifx\csname citenamefont\endcsname\relax
  \def\citenamefont#1{#1}\fi
\expandafter\ifx\csname url\endcsname\relax
  \def\url#1{\texttt{#1}}\fi
\expandafter\ifx\csname urlprefix\endcsname\relax\def\urlprefix{URL }\fi
\providecommand{\bibinfo}[2]{#2}
\providecommand{\eprint}[2][]{\url{#2}}

\bibitem[{\citenamefont{Alicea}(2012)}]{0034-4885-75-7-076501}
\bibinfo{author}{\bibfnamefont{J.}~\bibnamefont{Alicea}},
  \bibinfo{journal}{Reports on Progress in Physics}
  \textbf{\bibinfo{volume}{75}}, \bibinfo{pages}{076501}
  (\bibinfo{year}{2012}).

\bibitem[{\citenamefont{Beenakker}(2015)}]{RevModPhys.87.1037}
\bibinfo{author}{\bibfnamefont{C.~W.~J.} \bibnamefont{Beenakker}},
  \bibinfo{journal}{Rev. Mod. Phys.} \textbf{\bibinfo{volume}{87}},
  \bibinfo{pages}{1037} (\bibinfo{year}{2015}).

\bibitem[{\citenamefont{Leijnse and Flensberg}(2012)}]{0268-1242-27-12-124003}
\bibinfo{author}{\bibfnamefont{M.}~\bibnamefont{Leijnse}} \bibnamefont{and}
  \bibinfo{author}{\bibfnamefont{K.}~\bibnamefont{Flensberg}},
  \bibinfo{journal}{Semiconductor Science and Technology}
  \textbf{\bibinfo{volume}{27}}, \bibinfo{pages}{124003}
  (\bibinfo{year}{2012}).

\bibitem[{\citenamefont{Beenakker}(2013)}]{doi:10.1146/annurev-conmatphys-030212-184337}
\bibinfo{author}{\bibfnamefont{C.}~\bibnamefont{Beenakker}},
  \bibinfo{journal}{Annual Review of Condensed Matter Physics}
  \textbf{\bibinfo{volume}{4}}, \bibinfo{pages}{113} (\bibinfo{year}{2013}).

\bibitem[{\citenamefont{Kitaev}(2001)}]{1063-7869-44-10S-S29}
\bibinfo{author}{\bibfnamefont{A.~Y.} \bibnamefont{Kitaev}},
  \bibinfo{journal}{Physics-Uspekhi} \textbf{\bibinfo{volume}{44}},
  \bibinfo{pages}{131} (\bibinfo{year}{2001}).

\bibitem[{\citenamefont{Nayak et~al.}(2008)\citenamefont{Nayak, Simon, Stern,
  Freedman, and Das~Sarma}}]{RevModPhys.80.1083}
\bibinfo{author}{\bibfnamefont{C.}~\bibnamefont{Nayak}},
  \bibinfo{author}{\bibfnamefont{S.~H.} \bibnamefont{Simon}},
  \bibinfo{author}{\bibfnamefont{A.}~\bibnamefont{Stern}},
  \bibinfo{author}{\bibfnamefont{M.}~\bibnamefont{Freedman}}, \bibnamefont{and}
  \bibinfo{author}{\bibfnamefont{S.}~\bibnamefont{Das~Sarma}},
  \bibinfo{journal}{Rev. Mod. Phys.} \textbf{\bibinfo{volume}{80}},
  \bibinfo{pages}{1083} (\bibinfo{year}{2008}).

\bibitem[{\citenamefont{Alicea et~al.}(2011)\citenamefont{Alicea, Oreg, Refael,
  von Oppen, and Fisher}}]{ISI:000290150300017}
\bibinfo{author}{\bibfnamefont{J.}~\bibnamefont{Alicea}},
  \bibinfo{author}{\bibfnamefont{Y.}~\bibnamefont{Oreg}},
  \bibinfo{author}{\bibfnamefont{G.}~\bibnamefont{Refael}},
  \bibinfo{author}{\bibfnamefont{F.}~\bibnamefont{von Oppen}},
  \bibnamefont{and} \bibinfo{author}{\bibfnamefont{M.~P.~A.}
  \bibnamefont{Fisher}}, \bibinfo{journal}{NATURE PHYSICS}
  \textbf{\bibinfo{volume}{7}}, \bibinfo{pages}{412} (\bibinfo{year}{2011}).

\bibitem[{\citenamefont{Lutchyn et~al.}(2010)\citenamefont{Lutchyn, Sau, and
  Das~Sarma}}]{PhysRevLett.105.077001}
\bibinfo{author}{\bibfnamefont{R.~M.} \bibnamefont{Lutchyn}},
  \bibinfo{author}{\bibfnamefont{J.~D.} \bibnamefont{Sau}}, \bibnamefont{and}
  \bibinfo{author}{\bibfnamefont{S.}~\bibnamefont{Das~Sarma}},
  \bibinfo{journal}{Phys. Rev. Lett.} \textbf{\bibinfo{volume}{105}},
  \bibinfo{pages}{077001} (\bibinfo{year}{2010}).

\bibitem[{\citenamefont{Oreg et~al.}(2010)\citenamefont{Oreg, Refael, and von
  Oppen}}]{PhysRevLett.105.177002}
\bibinfo{author}{\bibfnamefont{Y.}~\bibnamefont{Oreg}},
  \bibinfo{author}{\bibfnamefont{G.}~\bibnamefont{Refael}}, \bibnamefont{and}
  \bibinfo{author}{\bibfnamefont{F.}~\bibnamefont{von Oppen}},
  \bibinfo{journal}{Phys. Rev. Lett.} \textbf{\bibinfo{volume}{105}},
  \bibinfo{pages}{177002} (\bibinfo{year}{2010}).

\bibitem[{\citenamefont{Pientka et~al.}(2015)\citenamefont{Pientka, Peng,
  Glazman, and von Oppen}}]{1402-4896-2015-T164-014008}
\bibinfo{author}{\bibfnamefont{F.}~\bibnamefont{Pientka}},
  \bibinfo{author}{\bibfnamefont{Y.}~\bibnamefont{Peng}},
  \bibinfo{author}{\bibfnamefont{L.}~\bibnamefont{Glazman}}, \bibnamefont{and}
  \bibinfo{author}{\bibfnamefont{F.}~\bibnamefont{von Oppen}},
  \bibinfo{journal}{Physica Scripta} \textbf{\bibinfo{volume}{2015}},
  \bibinfo{pages}{014008} (\bibinfo{year}{2015}).

\bibitem[{\citenamefont{Nadj-Perge et~al.}(2013)\citenamefont{Nadj-Perge,
  Drozdov, Bernevig, and Yazdani}}]{PhysRevB.88.020407}
\bibinfo{author}{\bibfnamefont{S.}~\bibnamefont{Nadj-Perge}},
  \bibinfo{author}{\bibfnamefont{I.~K.} \bibnamefont{Drozdov}},
  \bibinfo{author}{\bibfnamefont{B.~A.} \bibnamefont{Bernevig}},
  \bibnamefont{and} \bibinfo{author}{\bibfnamefont{A.}~\bibnamefont{Yazdani}},
  \bibinfo{journal}{Phys. Rev. B} \textbf{\bibinfo{volume}{88}},
  \bibinfo{pages}{020407} (\bibinfo{year}{2013}).

\bibitem[{\citenamefont{Nadj-Perge et~al.}(2014)\citenamefont{Nadj-Perge,
  Drozdov, Li, Chen, Jeon, Seo, MacDonald, Bernevig, and
  Yazdani}}]{Nadj-Perge602}
\bibinfo{author}{\bibfnamefont{S.}~\bibnamefont{Nadj-Perge}},
  \bibinfo{author}{\bibfnamefont{I.~K.} \bibnamefont{Drozdov}},
  \bibinfo{author}{\bibfnamefont{J.}~\bibnamefont{Li}},
  \bibinfo{author}{\bibfnamefont{H.}~\bibnamefont{Chen}},
  \bibinfo{author}{\bibfnamefont{S.}~\bibnamefont{Jeon}},
  \bibinfo{author}{\bibfnamefont{J.}~\bibnamefont{Seo}},
  \bibinfo{author}{\bibfnamefont{A.~H.} \bibnamefont{MacDonald}},
  \bibinfo{author}{\bibfnamefont{B.~A.} \bibnamefont{Bernevig}},
  \bibnamefont{and} \bibinfo{author}{\bibfnamefont{A.}~\bibnamefont{Yazdani}},
  \bibinfo{journal}{Science} \textbf{\bibinfo{volume}{346}},
  \bibinfo{pages}{602} (\bibinfo{year}{2014}).

\bibitem[{\citenamefont{Churchill et~al.}(2013)\citenamefont{Churchill, Fatemi,
  Grove-Rasmussen, Deng, Caroff, Xu, and Marcus}}]{PhysRevB.87.241401}
\bibinfo{author}{\bibfnamefont{H.~O.~H.} \bibnamefont{Churchill}},
  \bibinfo{author}{\bibfnamefont{V.}~\bibnamefont{Fatemi}},
  \bibinfo{author}{\bibfnamefont{K.}~\bibnamefont{Grove-Rasmussen}},
  \bibinfo{author}{\bibfnamefont{M.~T.} \bibnamefont{Deng}},
  \bibinfo{author}{\bibfnamefont{P.}~\bibnamefont{Caroff}},
  \bibinfo{author}{\bibfnamefont{H.~Q.} \bibnamefont{Xu}}, \bibnamefont{and}
  \bibinfo{author}{\bibfnamefont{C.~M.} \bibnamefont{Marcus}},
  \bibinfo{journal}{Phys. Rev. B} \textbf{\bibinfo{volume}{87}},
  \bibinfo{pages}{241401} (\bibinfo{year}{2013}).

\bibitem[{\citenamefont{Finck et~al.}(2013)\citenamefont{Finck, Van~Harlingen,
  Mohseni, Jung, and Li}}]{PhysRevLett.110.126406}
\bibinfo{author}{\bibfnamefont{A.~D.~K.} \bibnamefont{Finck}},
  \bibinfo{author}{\bibfnamefont{D.~J.} \bibnamefont{Van~Harlingen}},
  \bibinfo{author}{\bibfnamefont{P.~K.} \bibnamefont{Mohseni}},
  \bibinfo{author}{\bibfnamefont{K.}~\bibnamefont{Jung}}, \bibnamefont{and}
  \bibinfo{author}{\bibfnamefont{X.}~\bibnamefont{Li}}, \bibinfo{journal}{Phys.
  Rev. Lett.} \textbf{\bibinfo{volume}{110}}, \bibinfo{pages}{126406}
  (\bibinfo{year}{2013}).

\bibitem[{\citenamefont{Mourik et~al.}(2012)\citenamefont{Mourik, Zuo, Frolov,
  Plissard, Bakkers, and Kouwenhoven}}]{Mourik1003}
\bibinfo{author}{\bibfnamefont{V.}~\bibnamefont{Mourik}},
  \bibinfo{author}{\bibfnamefont{K.}~\bibnamefont{Zuo}},
  \bibinfo{author}{\bibfnamefont{S.~M.} \bibnamefont{Frolov}},
  \bibinfo{author}{\bibfnamefont{S.~R.} \bibnamefont{Plissard}},
  \bibinfo{author}{\bibfnamefont{E.~P. A.~M.} \bibnamefont{Bakkers}},
  \bibnamefont{and} \bibinfo{author}{\bibfnamefont{L.~P.}
  \bibnamefont{Kouwenhoven}}, \bibinfo{journal}{Science}
  \textbf{\bibinfo{volume}{336}}, \bibinfo{pages}{1003} (\bibinfo{year}{2012}).

\bibitem[{\citenamefont{Deng et~al.}(2012)\citenamefont{Deng, Yu, Huang,
  Larsson, Caroff, and Xu}}]{doi:10.1021/nl303758w}
\bibinfo{author}{\bibfnamefont{M.~T.} \bibnamefont{Deng}},
  \bibinfo{author}{\bibfnamefont{C.~L.} \bibnamefont{Yu}},
  \bibinfo{author}{\bibfnamefont{G.~Y.} \bibnamefont{Huang}},
  \bibinfo{author}{\bibfnamefont{M.}~\bibnamefont{Larsson}},
  \bibinfo{author}{\bibfnamefont{P.}~\bibnamefont{Caroff}}, \bibnamefont{and}
  \bibinfo{author}{\bibfnamefont{H.~Q.} \bibnamefont{Xu}},
  \bibinfo{journal}{Nano Letters} \textbf{\bibinfo{volume}{12}},
  \bibinfo{pages}{6414} (\bibinfo{year}{2012}), \bibinfo{note}{pMID: 23181691}.

\bibitem[{\citenamefont{Rokhinson et~al.}(2012)\citenamefont{Rokhinson, Liu,
  and Furdyna}}]{ISI:000310836700016}
\bibinfo{author}{\bibfnamefont{L.~P.} \bibnamefont{Rokhinson}},
  \bibinfo{author}{\bibfnamefont{X.}~\bibnamefont{Liu}}, \bibnamefont{and}
  \bibinfo{author}{\bibfnamefont{J.~K.} \bibnamefont{Furdyna}},
  \bibinfo{journal}{NATURE PHYSICS} \textbf{\bibinfo{volume}{8}},
  \bibinfo{pages}{795} (\bibinfo{year}{2012}).

\bibitem[{\citenamefont{Deng et~al.}(2014)\citenamefont{Deng, Yu, Huang,
  Larsson, Caroff, and Xu}}]{ISI:000346257700009}
\bibinfo{author}{\bibfnamefont{M.~T.} \bibnamefont{Deng}},
  \bibinfo{author}{\bibfnamefont{C.~L.} \bibnamefont{Yu}},
  \bibinfo{author}{\bibfnamefont{G.~Y.} \bibnamefont{Huang}},
  \bibinfo{author}{\bibfnamefont{M.}~\bibnamefont{Larsson}},
  \bibinfo{author}{\bibfnamefont{P.}~\bibnamefont{Caroff}}, \bibnamefont{and}
  \bibinfo{author}{\bibfnamefont{H.~Q.} \bibnamefont{Xu}},
  \bibinfo{journal}{SCIENTIFIC REPORTS} \textbf{\bibinfo{volume}{4}}
  (\bibinfo{year}{2014}).

\bibitem[{\citenamefont{Wiedenmann et~al.}(2016)\citenamefont{Wiedenmann,
  Bocquillon, Deacon, Hartinger, Herrmann, Klapwijk, Maier, Ames, Bruene, Gould
  et~al.}}]{ISI:000369019100004}
\bibinfo{author}{\bibfnamefont{J.}~\bibnamefont{Wiedenmann}},
  \bibinfo{author}{\bibfnamefont{E.}~\bibnamefont{Bocquillon}},
  \bibinfo{author}{\bibfnamefont{R.~S.} \bibnamefont{Deacon}},
  \bibinfo{author}{\bibfnamefont{S.}~\bibnamefont{Hartinger}},
  \bibinfo{author}{\bibfnamefont{O.}~\bibnamefont{Herrmann}},
  \bibinfo{author}{\bibfnamefont{T.~M.} \bibnamefont{Klapwijk}},
  \bibinfo{author}{\bibfnamefont{L.}~\bibnamefont{Maier}},
  \bibinfo{author}{\bibfnamefont{C.}~\bibnamefont{Ames}},
  \bibinfo{author}{\bibfnamefont{C.}~\bibnamefont{Bruene}},
  \bibinfo{author}{\bibfnamefont{C.}~\bibnamefont{Gould}},
  \bibnamefont{et~al.}, \bibinfo{journal}{NATURE COMMUNICATIONS}
  \textbf{\bibinfo{volume}{7}} (\bibinfo{year}{2016}).

\bibitem[{\citenamefont{Das et~al.}(2012)\citenamefont{Das, Ronen, Most, Oreg,
  Heiblum, and Shtrikman}}]{ISI:000311888200016}
\bibinfo{author}{\bibfnamefont{A.}~\bibnamefont{Das}},
  \bibinfo{author}{\bibfnamefont{Y.}~\bibnamefont{Ronen}},
  \bibinfo{author}{\bibfnamefont{Y.}~\bibnamefont{Most}},
  \bibinfo{author}{\bibfnamefont{Y.}~\bibnamefont{Oreg}},
  \bibinfo{author}{\bibfnamefont{M.}~\bibnamefont{Heiblum}}, \bibnamefont{and}
  \bibinfo{author}{\bibfnamefont{H.}~\bibnamefont{Shtrikman}},
  \bibinfo{journal}{NATURE PHYSICS} \textbf{\bibinfo{volume}{8}},
  \bibinfo{pages}{887} (\bibinfo{year}{2012}).

\bibitem[{\citenamefont{{Deacon} et~al.}(2016)\citenamefont{{Deacon},
  {Wiedenmann}, {Bocquillon}, {Dom{\'{\i}nguez}}, {Klapwijk}, {Leubner},
  {Br{\"u}ne}, {Hankiewicz}, {Tarucha}, {Ishibashi}
  et~al.}}]{2016arXiv160309611D}
\bibinfo{author}{\bibfnamefont{R.~S.} \bibnamefont{{Deacon}}},
  \bibinfo{author}{\bibfnamefont{J.}~\bibnamefont{{Wiedenmann}}},
  \bibinfo{author}{\bibfnamefont{E.}~\bibnamefont{{Bocquillon}}},
  \bibinfo{author}{\bibfnamefont{F.}~\bibnamefont{{Dom{\'{\i}nguez}}}},
  \bibinfo{author}{\bibfnamefont{T.~M.} \bibnamefont{{Klapwijk}}},
  \bibinfo{author}{\bibfnamefont{P.}~\bibnamefont{{Leubner}}},
  \bibinfo{author}{\bibfnamefont{C.}~\bibnamefont{{Br{\"u}ne}}},
  \bibinfo{author}{\bibfnamefont{E.~M.} \bibnamefont{{Hankiewicz}}},
  \bibinfo{author}{\bibfnamefont{S.}~\bibnamefont{{Tarucha}}},
  \bibinfo{author}{\bibfnamefont{K.}~\bibnamefont{{Ishibashi}}},
  \bibnamefont{et~al.}, \bibinfo{journal}{ArXiv e-prints}
  (\bibinfo{year}{2016}), \eprint{1603.09611}.

\bibitem[{\citenamefont{Hui et~al.}(2014)\citenamefont{Hui, Sau, and
  Das~Sarma}}]{PhysRevB.90.174206}
\bibinfo{author}{\bibfnamefont{H.-Y.} \bibnamefont{Hui}},
  \bibinfo{author}{\bibfnamefont{J.~D.} \bibnamefont{Sau}}, \bibnamefont{and}
  \bibinfo{author}{\bibfnamefont{S.}~\bibnamefont{Das~Sarma}},
  \bibinfo{journal}{Phys. Rev. B} \textbf{\bibinfo{volume}{90}},
  \bibinfo{pages}{174206} (\bibinfo{year}{2014}).

\bibitem[{\citenamefont{Sau and Das~Sarma}(2013)}]{PhysRevB.88.064506}
\bibinfo{author}{\bibfnamefont{J.~D.} \bibnamefont{Sau}} \bibnamefont{and}
  \bibinfo{author}{\bibfnamefont{S.}~\bibnamefont{Das~Sarma}},
  \bibinfo{journal}{Phys. Rev. B} \textbf{\bibinfo{volume}{88}},
  \bibinfo{pages}{064506} (\bibinfo{year}{2013}).

\bibitem[{\citenamefont{Lobos et~al.}(2012)\citenamefont{Lobos, Lutchyn, and
  Das~Sarma}}]{PhysRevLett.109.146403}
\bibinfo{author}{\bibfnamefont{A.~M.} \bibnamefont{Lobos}},
  \bibinfo{author}{\bibfnamefont{R.~M.} \bibnamefont{Lutchyn}},
  \bibnamefont{and}
  \bibinfo{author}{\bibfnamefont{S.}~\bibnamefont{Das~Sarma}},
  \bibinfo{journal}{Phys. Rev. Lett.} \textbf{\bibinfo{volume}{109}},
  \bibinfo{pages}{146403} (\bibinfo{year}{2012}).

\bibitem[{\citenamefont{Neven et~al.}(2013)\citenamefont{Neven, Bagrets, and
  Altland}}]{1367-2630-15-5-055019}
\bibinfo{author}{\bibfnamefont{P.}~\bibnamefont{Neven}},
  \bibinfo{author}{\bibfnamefont{D.}~\bibnamefont{Bagrets}}, \bibnamefont{and}
  \bibinfo{author}{\bibfnamefont{A.}~\bibnamefont{Altland}},
  \bibinfo{journal}{New Journal of Physics} \textbf{\bibinfo{volume}{15}},
  \bibinfo{pages}{055019} (\bibinfo{year}{2013}).

\bibitem[{\citenamefont{Rainis et~al.}(2013)\citenamefont{Rainis, Trifunovic,
  Klinovaja, and Loss}}]{PhysRevB.87.024515}
\bibinfo{author}{\bibfnamefont{D.}~\bibnamefont{Rainis}},
  \bibinfo{author}{\bibfnamefont{L.}~\bibnamefont{Trifunovic}},
  \bibinfo{author}{\bibfnamefont{J.}~\bibnamefont{Klinovaja}},
  \bibnamefont{and} \bibinfo{author}{\bibfnamefont{D.}~\bibnamefont{Loss}},
  \bibinfo{journal}{Phys. Rev. B} \textbf{\bibinfo{volume}{87}},
  \bibinfo{pages}{024515} (\bibinfo{year}{2013}).

\bibitem[{\citenamefont{Pikulin et~al.}(2012)\citenamefont{Pikulin, Dahlhaus,
  Wimmer, Schomerus, and Beenakker}}]{1367-2630-14-12-125011}
\bibinfo{author}{\bibfnamefont{D.~I.} \bibnamefont{Pikulin}},
  \bibinfo{author}{\bibfnamefont{J.~P.} \bibnamefont{Dahlhaus}},
  \bibinfo{author}{\bibfnamefont{M.}~\bibnamefont{Wimmer}},
  \bibinfo{author}{\bibfnamefont{H.}~\bibnamefont{Schomerus}},
  \bibnamefont{and} \bibinfo{author}{\bibfnamefont{C.~W.~J.}
  \bibnamefont{Beenakker}}, \bibinfo{journal}{New Journal of Physics}
  \textbf{\bibinfo{volume}{14}}, \bibinfo{pages}{125011}
  (\bibinfo{year}{2012}).

\bibitem[{\citenamefont{Liu et~al.}(2012)\citenamefont{Liu, Potter, Law, and
  Lee}}]{PhysRevLett.109.267002}
\bibinfo{author}{\bibfnamefont{J.}~\bibnamefont{Liu}},
  \bibinfo{author}{\bibfnamefont{A.~C.} \bibnamefont{Potter}},
  \bibinfo{author}{\bibfnamefont{K.~T.} \bibnamefont{Law}}, \bibnamefont{and}
  \bibinfo{author}{\bibfnamefont{P.~A.} \bibnamefont{Lee}},
  \bibinfo{journal}{Phys. Rev. Lett.} \textbf{\bibinfo{volume}{109}},
  \bibinfo{pages}{267002} (\bibinfo{year}{2012}).

\bibitem[{\citenamefont{Bagrets and Altland}(2012)}]{PhysRevLett.109.227005}
\bibinfo{author}{\bibfnamefont{D.}~\bibnamefont{Bagrets}} \bibnamefont{and}
  \bibinfo{author}{\bibfnamefont{A.}~\bibnamefont{Altland}},
  \bibinfo{journal}{Phys. Rev. Lett.} \textbf{\bibinfo{volume}{109}},
  \bibinfo{pages}{227005} (\bibinfo{year}{2012}).

\bibitem[{\citenamefont{Roy et~al.}(2013)\citenamefont{Roy, Bondyopadhaya, and
  Tewari}}]{PhysRevB.88.020502}
\bibinfo{author}{\bibfnamefont{D.}~\bibnamefont{Roy}},
  \bibinfo{author}{\bibfnamefont{N.}~\bibnamefont{Bondyopadhaya}},
  \bibnamefont{and} \bibinfo{author}{\bibfnamefont{S.}~\bibnamefont{Tewari}},
  \bibinfo{journal}{Phys. Rev. B} \textbf{\bibinfo{volume}{88}},
  \bibinfo{pages}{020502} (\bibinfo{year}{2013}).

\bibitem[{\citenamefont{Stanescu and Tewari}(2014)}]{PhysRevB.89.220507}
\bibinfo{author}{\bibfnamefont{T.~D.} \bibnamefont{Stanescu}} \bibnamefont{and}
  \bibinfo{author}{\bibfnamefont{S.}~\bibnamefont{Tewari}},
  \bibinfo{journal}{Phys. Rev. B} \textbf{\bibinfo{volume}{89}},
  \bibinfo{pages}{220507} (\bibinfo{year}{2014}).

\bibitem[{\citenamefont{Kells et~al.}(2012)\citenamefont{Kells, Meidan, and
  Brouwer}}]{PhysRevB.86.100503}
\bibinfo{author}{\bibfnamefont{G.}~\bibnamefont{Kells}},
  \bibinfo{author}{\bibfnamefont{D.}~\bibnamefont{Meidan}}, \bibnamefont{and}
  \bibinfo{author}{\bibfnamefont{P.~W.} \bibnamefont{Brouwer}},
  \bibinfo{journal}{Phys. Rev. B} \textbf{\bibinfo{volume}{86}},
  \bibinfo{pages}{100503} (\bibinfo{year}{2012}).

\bibitem[{\citenamefont{Lee et~al.}(2012)\citenamefont{Lee, Jiang, Aguado,
  Katsaros, Lieber, and De~Franceschi}}]{PhysRevLett.109.186802}
\bibinfo{author}{\bibfnamefont{E.~J.~H.} \bibnamefont{Lee}},
  \bibinfo{author}{\bibfnamefont{X.}~\bibnamefont{Jiang}},
  \bibinfo{author}{\bibfnamefont{R.}~\bibnamefont{Aguado}},
  \bibinfo{author}{\bibfnamefont{G.}~\bibnamefont{Katsaros}},
  \bibinfo{author}{\bibfnamefont{C.~M.} \bibnamefont{Lieber}},
  \bibnamefont{and}
  \bibinfo{author}{\bibfnamefont{S.}~\bibnamefont{De~Franceschi}},
  \bibinfo{journal}{Phys. Rev. Lett.} \textbf{\bibinfo{volume}{109}},
  \bibinfo{pages}{186802} (\bibinfo{year}{2012}).

\bibitem[{\citenamefont{Lee et~al.}(2014)\citenamefont{Lee, Jiang, Houzet,
  Aguado, Lieber, and De~Franceschi}}]{ISI:000329315000020}
\bibinfo{author}{\bibfnamefont{E.~J.~H.} \bibnamefont{Lee}},
  \bibinfo{author}{\bibfnamefont{X.}~\bibnamefont{Jiang}},
  \bibinfo{author}{\bibfnamefont{M.}~\bibnamefont{Houzet}},
  \bibinfo{author}{\bibfnamefont{R.}~\bibnamefont{Aguado}},
  \bibinfo{author}{\bibfnamefont{C.~M.} \bibnamefont{Lieber}},
  \bibnamefont{and}
  \bibinfo{author}{\bibfnamefont{S.}~\bibnamefont{De~Franceschi}},
  \bibinfo{journal}{NATURE NANOTECHNOLOGY} \textbf{\bibinfo{volume}{9}},
  \bibinfo{pages}{79} (\bibinfo{year}{2014}), ISSN \bibinfo{issn}{1748-3387}.

\bibitem[{\citenamefont{{Moore} et~al.}(2017)\citenamefont{{Moore}, {Stanescu},
  and {Tewari}}}]{2017arXiv171106256M}
\bibinfo{author}{\bibfnamefont{C.}~\bibnamefont{{Moore}}},
  \bibinfo{author}{\bibfnamefont{T.~D.} \bibnamefont{{Stanescu}}},
  \bibnamefont{and} \bibinfo{author}{\bibfnamefont{S.}~\bibnamefont{{Tewari}}},
  \bibinfo{journal}{ArXiv e-prints}  (\bibinfo{year}{2017}),
  \eprint{1711.06256}.

\bibitem[{\citenamefont{Kwon et~al.}(2004{\natexlab{a}})\citenamefont{Kwon,
  Sengupta, and Yakovenko}}]{Kwon2004}
\bibinfo{author}{\bibfnamefont{H.-J.} \bibnamefont{Kwon}},
  \bibinfo{author}{\bibfnamefont{K.}~\bibnamefont{Sengupta}}, \bibnamefont{and}
  \bibinfo{author}{\bibfnamefont{V.~M.} \bibnamefont{Yakovenko}},
  \bibinfo{journal}{The European Physical Journal B - Condensed Matter and
  Complex Systems} \textbf{\bibinfo{volume}{37}}, \bibinfo{pages}{349}
  (\bibinfo{year}{2004}{\natexlab{a}}).

\bibitem[{\citenamefont{Fu and Kane}(2009)}]{PhysRevB.79.161408}
\bibinfo{author}{\bibfnamefont{L.}~\bibnamefont{Fu}} \bibnamefont{and}
  \bibinfo{author}{\bibfnamefont{C.~L.} \bibnamefont{Kane}},
  \bibinfo{journal}{Phys. Rev. B} \textbf{\bibinfo{volume}{79}},
  \bibinfo{pages}{161408} (\bibinfo{year}{2009}).

\bibitem[{\citenamefont{Kwon et~al.}(2004{\natexlab{b}})\citenamefont{Kwon,
  Yakovenko, and Sengupta}}]{:/content/aip/journal/ltp/30/7/10.1063/1.1789931}
\bibinfo{author}{\bibfnamefont{H.-J.} \bibnamefont{Kwon}},
  \bibinfo{author}{\bibfnamefont{V.~M.} \bibnamefont{Yakovenko}},
  \bibnamefont{and} \bibinfo{author}{\bibfnamefont{K.}~\bibnamefont{Sengupta}},
  \bibinfo{journal}{Low Temperature Physics} \textbf{\bibinfo{volume}{30}},
  \bibinfo{pages}{613} (\bibinfo{year}{2004}{\natexlab{b}}).

\bibitem[{\citenamefont{Badiane et~al.}(2013)\citenamefont{Badiane, Glazman,
  Houzet, and Meyer}}]{Badiane2013840}
\bibinfo{author}{\bibfnamefont{D.~M.} \bibnamefont{Badiane}},
  \bibinfo{author}{\bibfnamefont{L.~I.} \bibnamefont{Glazman}},
  \bibinfo{author}{\bibfnamefont{M.}~\bibnamefont{Houzet}}, \bibnamefont{and}
  \bibinfo{author}{\bibfnamefont{J.~S.} \bibnamefont{Meyer}},
  \bibinfo{journal}{Comptes Rendus Physique} \textbf{\bibinfo{volume}{14}},
  \bibinfo{pages}{840 } (\bibinfo{year}{2013}).

\bibitem[{\citenamefont{{Sau} et~al.}(2012)\citenamefont{{Sau}, {Berg}, and
  {Halperin}}}]{2012arXiv1206.4596S}
\bibinfo{author}{\bibfnamefont{J.~D.} \bibnamefont{{Sau}}},
  \bibinfo{author}{\bibfnamefont{E.}~\bibnamefont{{Berg}}}, \bibnamefont{and}
  \bibinfo{author}{\bibfnamefont{B.~I.} \bibnamefont{{Halperin}}},
  \bibinfo{journal}{ArXiv e-prints}  (\bibinfo{year}{2012}),
  \eprint{1206.4596}.

\bibitem[{\citenamefont{Sau and Setiawan}(2017)}]{PhysRevB.95.060501}
\bibinfo{author}{\bibfnamefont{J.~D.} \bibnamefont{Sau}} \bibnamefont{and}
  \bibinfo{author}{\bibfnamefont{F.}~\bibnamefont{Setiawan}},
  \bibinfo{journal}{Phys. Rev. B} \textbf{\bibinfo{volume}{95}},
  \bibinfo{pages}{060501} (\bibinfo{year}{2017}).

\bibitem[{\citenamefont{Fazio and van~der Zant}(2001)}]{Fazio2001235}
\bibinfo{author}{\bibfnamefont{R.}~\bibnamefont{Fazio}} \bibnamefont{and}
  \bibinfo{author}{\bibfnamefont{H.}~\bibnamefont{van~der Zant}},
  \bibinfo{journal}{Physics Reports} \textbf{\bibinfo{volume}{355}},
  \bibinfo{pages}{235 } (\bibinfo{year}{2001}).

\bibitem[{\citenamefont{Pop et~al.}(2010)\citenamefont{Pop, Protopopov, Lecocq,
  Peng, Pannetier, Buisson, and Guichard}}]{ISI:000280559300014}
\bibinfo{author}{\bibfnamefont{I.~M.} \bibnamefont{Pop}},
  \bibinfo{author}{\bibfnamefont{I.}~\bibnamefont{Protopopov}},
  \bibinfo{author}{\bibfnamefont{F.}~\bibnamefont{Lecocq}},
  \bibinfo{author}{\bibfnamefont{Z.}~\bibnamefont{Peng}},
  \bibinfo{author}{\bibfnamefont{B.}~\bibnamefont{Pannetier}},
  \bibinfo{author}{\bibfnamefont{O.}~\bibnamefont{Buisson}}, \bibnamefont{and}
  \bibinfo{author}{\bibfnamefont{W.}~\bibnamefont{Guichard}},
  \bibinfo{journal}{NATURE PHYSICS} \textbf{\bibinfo{volume}{6}},
  \bibinfo{pages}{589} (\bibinfo{year}{2010}).

\bibitem[{\citenamefont{Haviland and Delsing}(1996)}]{PhysRevB.54.R6857}
\bibinfo{author}{\bibfnamefont{D.~B.} \bibnamefont{Haviland}} \bibnamefont{and}
  \bibinfo{author}{\bibfnamefont{P.}~\bibnamefont{Delsing}},
  \bibinfo{journal}{Phys. Rev. B} \textbf{\bibinfo{volume}{54}},
  \bibinfo{pages}{R6857} (\bibinfo{year}{1996}).

\bibitem[{\citenamefont{Chow et~al.}(1998)\citenamefont{Chow, Delsing, and
  Haviland}}]{PhysRevLett.81.204}
\bibinfo{author}{\bibfnamefont{E.}~\bibnamefont{Chow}},
  \bibinfo{author}{\bibfnamefont{P.}~\bibnamefont{Delsing}}, \bibnamefont{and}
  \bibinfo{author}{\bibfnamefont{D.~B.} \bibnamefont{Haviland}},
  \bibinfo{journal}{Phys. Rev. Lett.} \textbf{\bibinfo{volume}{81}},
  \bibinfo{pages}{204} (\bibinfo{year}{1998}).

\bibitem[{\citenamefont{Haviland et~al.}(2000)\citenamefont{Haviland,
  Andersson, and {\AA}gren}}]{Haviland2000}
\bibinfo{author}{\bibfnamefont{D.~B.} \bibnamefont{Haviland}},
  \bibinfo{author}{\bibfnamefont{K.}~\bibnamefont{Andersson}},
  \bibnamefont{and}
  \bibinfo{author}{\bibfnamefont{P.}~\bibnamefont{{\AA}gren}},
  \bibinfo{journal}{Journal of Low Temperature Physics}
  \textbf{\bibinfo{volume}{118}}, \bibinfo{pages}{733} (\bibinfo{year}{2000}).

\bibitem[{\citenamefont{Rubbert and Akhmerov}(2016)}]{PhysRevB.94.115430}
\bibinfo{author}{\bibfnamefont{S.}~\bibnamefont{Rubbert}} \bibnamefont{and}
  \bibinfo{author}{\bibfnamefont{A.~R.} \bibnamefont{Akhmerov}},
  \bibinfo{journal}{Phys. Rev. B} \textbf{\bibinfo{volume}{94}},
  \bibinfo{pages}{115430} (\bibinfo{year}{2016}).

\bibitem[{\citenamefont{Hyart et~al.}(2013)\citenamefont{Hyart, van Heck,
  Fulga, Burrello, Akhmerov, and Beenakker}}]{PhysRevB.88.035121}
\bibinfo{author}{\bibfnamefont{T.}~\bibnamefont{Hyart}},
  \bibinfo{author}{\bibfnamefont{B.}~\bibnamefont{van Heck}},
  \bibinfo{author}{\bibfnamefont{I.~C.} \bibnamefont{Fulga}},
  \bibinfo{author}{\bibfnamefont{M.}~\bibnamefont{Burrello}},
  \bibinfo{author}{\bibfnamefont{A.~R.} \bibnamefont{Akhmerov}},
  \bibnamefont{and} \bibinfo{author}{\bibfnamefont{C.~W.~J.}
  \bibnamefont{Beenakker}}, \bibinfo{journal}{Phys. Rev. B}
  \textbf{\bibinfo{volume}{88}}, \bibinfo{pages}{035121}
  (\bibinfo{year}{2013}).

\bibitem[{\citenamefont{Flensberg}(2011)}]{PhysRevLett.106.090503}
\bibinfo{author}{\bibfnamefont{K.}~\bibnamefont{Flensberg}},
  \bibinfo{journal}{Phys. Rev. Lett.} \textbf{\bibinfo{volume}{106}},
  \bibinfo{pages}{090503} (\bibinfo{year}{2011}).

\bibitem[{\citenamefont{Bonderson and Lutchyn}(2011)}]{PhysRevLett.106.130505}
\bibinfo{author}{\bibfnamefont{P.}~\bibnamefont{Bonderson}} \bibnamefont{and}
  \bibinfo{author}{\bibfnamefont{R.~M.} \bibnamefont{Lutchyn}},
  \bibinfo{journal}{Phys. Rev. Lett.} \textbf{\bibinfo{volume}{106}},
  \bibinfo{pages}{130505} (\bibinfo{year}{2011}).

\bibitem[{\citenamefont{van Heck et~al.}(2012)\citenamefont{van Heck, Akhmerov,
  Hassler, Burrello, and Beenakker}}]{1367-2630-14-3-035019}
\bibinfo{author}{\bibfnamefont{B.}~\bibnamefont{van Heck}},
  \bibinfo{author}{\bibfnamefont{A.~R.} \bibnamefont{Akhmerov}},
  \bibinfo{author}{\bibfnamefont{F.}~\bibnamefont{Hassler}},
  \bibinfo{author}{\bibfnamefont{M.}~\bibnamefont{Burrello}}, \bibnamefont{and}
  \bibinfo{author}{\bibfnamefont{C.~W.~J.} \bibnamefont{Beenakker}},
  \bibinfo{journal}{New Journal of Physics} \textbf{\bibinfo{volume}{14}},
  \bibinfo{pages}{035019} (\bibinfo{year}{2012}).

\bibitem[{\citenamefont{Aasen et~al.}(2016)\citenamefont{Aasen, Hell, Mishmash,
  Higginbotham, Danon, Leijnse, Jespersen, Folk, Marcus, Flensberg
  et~al.}}]{PhysRevX.6.031016}
\bibinfo{author}{\bibfnamefont{D.}~\bibnamefont{Aasen}},
  \bibinfo{author}{\bibfnamefont{M.}~\bibnamefont{Hell}},
  \bibinfo{author}{\bibfnamefont{R.~V.} \bibnamefont{Mishmash}},
  \bibinfo{author}{\bibfnamefont{A.}~\bibnamefont{Higginbotham}},
  \bibinfo{author}{\bibfnamefont{J.}~\bibnamefont{Danon}},
  \bibinfo{author}{\bibfnamefont{M.}~\bibnamefont{Leijnse}},
  \bibinfo{author}{\bibfnamefont{T.~S.} \bibnamefont{Jespersen}},
  \bibinfo{author}{\bibfnamefont{J.~A.} \bibnamefont{Folk}},
  \bibinfo{author}{\bibfnamefont{C.~M.} \bibnamefont{Marcus}},
  \bibinfo{author}{\bibfnamefont{K.}~\bibnamefont{Flensberg}},
  \bibnamefont{et~al.}, \bibinfo{journal}{Phys. Rev. X}
  \textbf{\bibinfo{volume}{6}}, \bibinfo{pages}{031016} (\bibinfo{year}{2016}).

\bibitem[{\citenamefont{Karzig et~al.}(2017)\citenamefont{Karzig, Knapp,
  Lutchyn, Bonderson, Hastings, Nayak, Alicea, Flensberg, Plugge, Oreg
  et~al.}}]{PhysRevB.95.235305}
\bibinfo{author}{\bibfnamefont{T.}~\bibnamefont{Karzig}},
  \bibinfo{author}{\bibfnamefont{C.}~\bibnamefont{Knapp}},
  \bibinfo{author}{\bibfnamefont{R.~M.} \bibnamefont{Lutchyn}},
  \bibinfo{author}{\bibfnamefont{P.}~\bibnamefont{Bonderson}},
  \bibinfo{author}{\bibfnamefont{M.~B.} \bibnamefont{Hastings}},
  \bibinfo{author}{\bibfnamefont{C.}~\bibnamefont{Nayak}},
  \bibinfo{author}{\bibfnamefont{J.}~\bibnamefont{Alicea}},
  \bibinfo{author}{\bibfnamefont{K.}~\bibnamefont{Flensberg}},
  \bibinfo{author}{\bibfnamefont{S.}~\bibnamefont{Plugge}},
  \bibinfo{author}{\bibfnamefont{Y.}~\bibnamefont{Oreg}}, \bibnamefont{et~al.},
  \bibinfo{journal}{Phys. Rev. B} \textbf{\bibinfo{volume}{95}},
  \bibinfo{pages}{235305} (\bibinfo{year}{2017}).

\bibitem[{\citenamefont{Gharavi et~al.}(2016)\citenamefont{Gharavi, Hoving, and
  Baugh}}]{PhysRevB.94.155417}
\bibinfo{author}{\bibfnamefont{K.}~\bibnamefont{Gharavi}},
  \bibinfo{author}{\bibfnamefont{D.}~\bibnamefont{Hoving}}, \bibnamefont{and}
  \bibinfo{author}{\bibfnamefont{J.}~\bibnamefont{Baugh}},
  \bibinfo{journal}{Phys. Rev. B} \textbf{\bibinfo{volume}{94}},
  \bibinfo{pages}{155417} (\bibinfo{year}{2016}).

\bibitem[{\citenamefont{Hoffman et~al.}(2016)\citenamefont{Hoffman, Schrade,
  Klinovaja, and Loss}}]{PhysRevB.94.045316}
\bibinfo{author}{\bibfnamefont{S.}~\bibnamefont{Hoffman}},
  \bibinfo{author}{\bibfnamefont{C.}~\bibnamefont{Schrade}},
  \bibinfo{author}{\bibfnamefont{J.}~\bibnamefont{Klinovaja}},
  \bibnamefont{and} \bibinfo{author}{\bibfnamefont{D.}~\bibnamefont{Loss}},
  \bibinfo{journal}{Phys. Rev. B} \textbf{\bibinfo{volume}{94}},
  \bibinfo{pages}{045316} (\bibinfo{year}{2016}).

\bibitem[{\citenamefont{Vernek et~al.}(2014)\citenamefont{Vernek, Penteado,
  Seridonio, and Egues}}]{PhysRevB.89.165314}
\bibinfo{author}{\bibfnamefont{E.}~\bibnamefont{Vernek}},
  \bibinfo{author}{\bibfnamefont{P.~H.} \bibnamefont{Penteado}},
  \bibinfo{author}{\bibfnamefont{A.~C.} \bibnamefont{Seridonio}},
  \bibnamefont{and} \bibinfo{author}{\bibfnamefont{J.~C.} \bibnamefont{Egues}},
  \bibinfo{journal}{Phys. Rev. B} \textbf{\bibinfo{volume}{89}},
  \bibinfo{pages}{165314} (\bibinfo{year}{2014}).

\bibitem[{\citenamefont{Deng et~al.}(2016)\citenamefont{Deng, Vaitiekenas,
  Hansen, Danon, Leijnse, Flensberg, Nyg{\r a}rd, Krogstrup, and
  Marcus}}]{Deng1557}
\bibinfo{author}{\bibfnamefont{M.~T.} \bibnamefont{Deng}},
  \bibinfo{author}{\bibfnamefont{S.}~\bibnamefont{Vaitiekenas}},
  \bibinfo{author}{\bibfnamefont{E.~B.} \bibnamefont{Hansen}},
  \bibinfo{author}{\bibfnamefont{J.}~\bibnamefont{Danon}},
  \bibinfo{author}{\bibfnamefont{M.}~\bibnamefont{Leijnse}},
  \bibinfo{author}{\bibfnamefont{K.}~\bibnamefont{Flensberg}},
  \bibinfo{author}{\bibfnamefont{J.}~\bibnamefont{Nyg{\r a}rd}},
  \bibinfo{author}{\bibfnamefont{P.}~\bibnamefont{Krogstrup}},
  \bibnamefont{and} \bibinfo{author}{\bibfnamefont{C.~M.}
  \bibnamefont{Marcus}}, \bibinfo{journal}{Science}
  \textbf{\bibinfo{volume}{354}}, \bibinfo{pages}{1557} (\bibinfo{year}{2016}),
  ISSN \bibinfo{issn}{0036-8075}.

\bibitem[{\citenamefont{Li et~al.}(2014)\citenamefont{Li, Yu, Lin, and
  You}}]{ISI:000335642000003}
\bibinfo{author}{\bibfnamefont{J.}~\bibnamefont{Li}},
  \bibinfo{author}{\bibfnamefont{T.}~\bibnamefont{Yu}},
  \bibinfo{author}{\bibfnamefont{H.-Q.} \bibnamefont{Lin}}, \bibnamefont{and}
  \bibinfo{author}{\bibfnamefont{J.~Q.} \bibnamefont{You}},
  \bibinfo{journal}{SCIENTIFIC REPORTS} \textbf{\bibinfo{volume}{4}}
  (\bibinfo{year}{2014}), ISSN \bibinfo{issn}{2045-2322}.

\bibitem[{\citenamefont{Plugge et~al.}(2017)\citenamefont{Plugge, Rasmussen,
  Egger, and Flensberg}}]{1367-2630-19-1-012001}
\bibinfo{author}{\bibfnamefont{S.}~\bibnamefont{Plugge}},
  \bibinfo{author}{\bibfnamefont{A.}~\bibnamefont{Rasmussen}},
  \bibinfo{author}{\bibfnamefont{R.}~\bibnamefont{Egger}}, \bibnamefont{and}
  \bibinfo{author}{\bibfnamefont{K.}~\bibnamefont{Flensberg}},
  \bibinfo{journal}{New Journal of Physics} \textbf{\bibinfo{volume}{19}},
  \bibinfo{pages}{012001} (\bibinfo{year}{2017}).

\bibitem[{\citenamefont{Liu and Baranger}(2011)}]{PhysRevB.84.201308}
\bibinfo{author}{\bibfnamefont{D.~E.} \bibnamefont{Liu}} \bibnamefont{and}
  \bibinfo{author}{\bibfnamefont{H.~U.} \bibnamefont{Baranger}},
  \bibinfo{journal}{Phys. Rev. B} \textbf{\bibinfo{volume}{84}},
  \bibinfo{pages}{201308} (\bibinfo{year}{2011}).

\bibitem[{\citenamefont{{Clarke}}(2017)}]{2017arXiv170201740C}
\bibinfo{author}{\bibfnamefont{D.~J.} \bibnamefont{{Clarke}}},
  \bibinfo{journal}{ArXiv e-prints}  (\bibinfo{year}{2017}),
  \eprint{1702.01740}.

\bibitem[{\citenamefont{Prada et~al.}(2017)\citenamefont{Prada, Aguado, and
  San-Jose}}]{PhysRevB.96.085418}
\bibinfo{author}{\bibfnamefont{E.}~\bibnamefont{Prada}},
  \bibinfo{author}{\bibfnamefont{R.}~\bibnamefont{Aguado}}, \bibnamefont{and}
  \bibinfo{author}{\bibfnamefont{P.}~\bibnamefont{San-Jose}},
  \bibinfo{journal}{Phys. Rev. B} \textbf{\bibinfo{volume}{96}},
  \bibinfo{pages}{085418} (\bibinfo{year}{2017}).

\bibitem[{\citenamefont{Feldman et~al.}(2016)\citenamefont{Feldman, Randeria,
  Li, Jeon, Xie, Wang, Drozdov, Bernevig, and Yazdani}}]{Feldman}
\bibinfo{author}{\bibfnamefont{B.~E.} \bibnamefont{Feldman}},
  \bibinfo{author}{\bibfnamefont{M.~T.} \bibnamefont{Randeria}},
  \bibinfo{author}{\bibfnamefont{J.}~\bibnamefont{Li}},
  \bibinfo{author}{\bibfnamefont{S.}~\bibnamefont{Jeon}},
  \bibinfo{author}{\bibfnamefont{Y.}~\bibnamefont{Xie}},
  \bibinfo{author}{\bibfnamefont{Z.}~\bibnamefont{Wang}},
  \bibinfo{author}{\bibfnamefont{I.~K.} \bibnamefont{Drozdov}},
  \bibinfo{author}{\bibfnamefont{B.~A.} \bibnamefont{Bernevig}},
  \bibnamefont{and} \bibinfo{author}{\bibfnamefont{A.}~\bibnamefont{Yazdani}},
  \bibinfo{journal}{Nat. Phys.} \textbf{\bibinfo{volume}{13}}
  (\bibinfo{year}{2016}).

\bibitem[{\citenamefont{Matveev et~al.}(2002)\citenamefont{Matveev, Larkin, and
  Glazman}}]{PhysRevLett.89.096802}
\bibinfo{author}{\bibfnamefont{K.~A.} \bibnamefont{Matveev}},
  \bibinfo{author}{\bibfnamefont{A.~I.} \bibnamefont{Larkin}},
  \bibnamefont{and} \bibinfo{author}{\bibfnamefont{L.~I.}
  \bibnamefont{Glazman}}, \bibinfo{journal}{Phys. Rev. Lett.}
  \textbf{\bibinfo{volume}{89}}, \bibinfo{pages}{096802}
  (\bibinfo{year}{2002}).

\bibitem[{\citenamefont{Rastelli et~al.}(2013)\citenamefont{Rastelli, Pop, and
  Hekking}}]{PhysRevB.87.174513}
\bibinfo{author}{\bibfnamefont{G.}~\bibnamefont{Rastelli}},
  \bibinfo{author}{\bibfnamefont{I.~M.} \bibnamefont{Pop}}, \bibnamefont{and}
  \bibinfo{author}{\bibfnamefont{F.~W.~J.} \bibnamefont{Hekking}},
  \bibinfo{journal}{Phys. Rev. B} \textbf{\bibinfo{volume}{87}},
  \bibinfo{pages}{174513} (\bibinfo{year}{2013}).

\bibitem[{\citenamefont{Fu}(2010)}]{PhysRevLett.104.056402}
\bibinfo{author}{\bibfnamefont{L.}~\bibnamefont{Fu}}, \bibinfo{journal}{Phys.
  Rev. Lett.} \textbf{\bibinfo{volume}{104}}, \bibinfo{pages}{056402}
  (\bibinfo{year}{2010}).

\bibitem[{\citenamefont{van Heck et~al.}(2011)\citenamefont{van Heck, Hassler,
  Akhmerov, and Beenakker}}]{PhysRevB.84.180502}
\bibinfo{author}{\bibfnamefont{B.}~\bibnamefont{van Heck}},
  \bibinfo{author}{\bibfnamefont{F.}~\bibnamefont{Hassler}},
  \bibinfo{author}{\bibfnamefont{A.~R.} \bibnamefont{Akhmerov}},
  \bibnamefont{and} \bibinfo{author}{\bibfnamefont{C.~W.~J.}
  \bibnamefont{Beenakker}}, \bibinfo{journal}{Phys. Rev. B}
  \textbf{\bibinfo{volume}{84}}, \bibinfo{pages}{180502}
  (\bibinfo{year}{2011}).

\bibitem[{\citenamefont{Pekker et~al.}(2013)\citenamefont{Pekker, Hou, Bergman,
  Goldberg, Adagideli, and Hassler}}]{PhysRevB.87.064506}
\bibinfo{author}{\bibfnamefont{D.}~\bibnamefont{Pekker}},
  \bibinfo{author}{\bibfnamefont{C.-Y.} \bibnamefont{Hou}},
  \bibinfo{author}{\bibfnamefont{D.~L.} \bibnamefont{Bergman}},
  \bibinfo{author}{\bibfnamefont{S.}~\bibnamefont{Goldberg}},
  \bibinfo{author}{\bibfnamefont{i.~d. I. m.~c.} \bibnamefont{Adagideli}},
  \bibnamefont{and} \bibinfo{author}{\bibfnamefont{F.}~\bibnamefont{Hassler}},
  \bibinfo{journal}{Phys. Rev. B} \textbf{\bibinfo{volume}{87}},
  \bibinfo{pages}{064506} (\bibinfo{year}{2013}).

\bibitem[{\citenamefont{L\'evy et~al.}(1990)\citenamefont{L\'evy, Dolan,
  Dunsmuir, and Bouchiat}}]{PhysRevLett.64.2074}
\bibinfo{author}{\bibfnamefont{L.~P.} \bibnamefont{L\'evy}},
  \bibinfo{author}{\bibfnamefont{G.}~\bibnamefont{Dolan}},
  \bibinfo{author}{\bibfnamefont{J.}~\bibnamefont{Dunsmuir}}, \bibnamefont{and}
  \bibinfo{author}{\bibfnamefont{H.}~\bibnamefont{Bouchiat}},
  \bibinfo{journal}{Phys. Rev. Lett.} \textbf{\bibinfo{volume}{64}},
  \bibinfo{pages}{2074} (\bibinfo{year}{1990}).

\bibitem[{\citenamefont{Bleszynski-Jayich
  et~al.}(2009)\citenamefont{Bleszynski-Jayich, Shanks, Peaudecerf, Ginossar,
  von Oppen, Glazman, and Harris}}]{Bleszynski-Jayich272}
\bibinfo{author}{\bibfnamefont{A.~C.} \bibnamefont{Bleszynski-Jayich}},
  \bibinfo{author}{\bibfnamefont{W.~E.} \bibnamefont{Shanks}},
  \bibinfo{author}{\bibfnamefont{B.}~\bibnamefont{Peaudecerf}},
  \bibinfo{author}{\bibfnamefont{E.}~\bibnamefont{Ginossar}},
  \bibinfo{author}{\bibfnamefont{F.}~\bibnamefont{von Oppen}},
  \bibinfo{author}{\bibfnamefont{L.}~\bibnamefont{Glazman}}, \bibnamefont{and}
  \bibinfo{author}{\bibfnamefont{J.~G.~E.} \bibnamefont{Harris}},
  \bibinfo{journal}{Science} \textbf{\bibinfo{volume}{326}},
  \bibinfo{pages}{272} (\bibinfo{year}{2009}), ISSN \bibinfo{issn}{0036-8075}.

\bibitem[{\citenamefont{Bary-Soroker et~al.}(2010)\citenamefont{Bary-Soroker,
  Entin-Wohlman, and Imry}}]{PhysRevB.82.144202}
\bibinfo{author}{\bibfnamefont{H.}~\bibnamefont{Bary-Soroker}},
  \bibinfo{author}{\bibfnamefont{O.}~\bibnamefont{Entin-Wohlman}},
  \bibnamefont{and} \bibinfo{author}{\bibfnamefont{Y.}~\bibnamefont{Imry}},
  \bibinfo{journal}{Phys. Rev. B} \textbf{\bibinfo{volume}{82}},
  \bibinfo{pages}{144202} (\bibinfo{year}{2010}).

\bibitem[{\citenamefont{Xu and Fu}(2010)}]{PhysRevB.81.134435}
\bibinfo{author}{\bibfnamefont{C.}~\bibnamefont{Xu}} \bibnamefont{and}
  \bibinfo{author}{\bibfnamefont{L.}~\bibnamefont{Fu}}, \bibinfo{journal}{Phys.
  Rev. B} \textbf{\bibinfo{volume}{81}}, \bibinfo{pages}{134435}
  (\bibinfo{year}{2010}).

\end{thebibliography}

\end{document}